\documentclass[11pt]{article}
\usepackage{graphics}
\usepackage{psfrag}
\usepackage{epsfig}
\usepackage{psfig}
\usepackage{epsf}
\usepackage{float}
\usepackage{latexsym}
\usepackage[german,USenglish]{babel}
\newcommand{\vs}{\vspace{-0.2cm}}
\newcommand{\beq}{\begin{equation}}
\newcommand{\eeq}{\end{equation}}
\newcommand{\beqa}{\begin{eqnarray}}
\newcommand{\eeqa}{\end{eqnarray}}
\newcommand{\nn}{\nonumber \\ }
\newcommand{\bma}{\begin{array}{cc}}
\newcommand{\ema}{\end{array}}

\newcommand{\rb}[1]{\raisebox{1.5ex}[-1.5ex]{#1}}

\def\3{{\ss}}

\def\vek #1 {\overrightarrow {#1}}
\newcommand{\fet}[1]{\mbox{\boldmath $#1$}}

\begin{document}

\hfill {\tiny FZJ-IKP(TH)-2002-03}

\begin{center}

{{\Large\bf Few--Nucleon Systems with Two--Nucleon Forces\\[0.3em]
from Chiral Effective Field Theory
}}

\end{center}

\vspace{.3in}

\begin{center}

{\large 
E. Epelbaum,$^\dagger$\footnote{email: 
                           evgeni.epelbaum@tp2.ruhr-uni-bochum.de}
A. Nogga,$^\star$\footnote{email:
                           anogga@physics.arizona.edu}
W. Gl\"ockle,$^\dagger$\footnote{email:
                           walter.gloeckle@tp2.ruhr-uni-bochum.de}
H. Kamada,$^\ast$\footnote{email:
                           kamada@mns.kyutech.ac.jp}
Ulf-G. Mei{\ss}ner,$^\ddagger$\footnote{email: 
                           u.meissner@fz-juelich.de}
H. Wita{\l}a$^\circ$}\footnote{email:
                           witala@if.uj.edu.pl}

\bigskip

$^\dagger${\it Ruhr-Universit\"at Bochum, Institut f{\"u}r
  Theoretische Physik II,\\ D-44870 Bochum, Germany}\\

\bigskip

$^\star${\it Department of  Physics, University of Arizona, Tucson, Arizona 85721, USA}\\

\bigskip
$^\ast${\it Department of  Physics, Faculty of Engineering, Kyushu Institute of Technology, \\
1-1 Sensuicho, Tobata, Kitakyushu 804-8550, Japan}\\

\bigskip

$^\ddagger${\it Forschungszentrum J\"ulich, Institut f\"ur Kernphysik 
(Theorie),\\ D-52425 J\"ulich, Germany}
\bigskip

$^\circ${\it  Jagiellonian University, Institut of Physics, Reymonta 4, \\
   30-059 Cracow, Poland}

\end{center}

\vspace{.3in}

\thispagestyle{empty}

\begin{abstract}

\noindent
Nucleon--nucleon (NN) forces  from chiral perturbation theory
at next--to--leading (NLO) and  next--to--next--to--leading order (NNLO) are
applied to systems with two, three and four nucleons. At NNLO, we consider
two versions of the chiral potential which differ in the 
strength of the two--pion--exchange (TPE)
but describe two nucleon observables equally well.
The NNLO potential leads to unphysical deeply bound
states in the low partial waves and 
effects of the 3N forces, which appear first at this order,
are expected to be large.
We provide arguments for
a reduction of the TPE potential and introduce the NNLO* 
version of the NN forces. 
We calculate $nd$ scattering observables as well as various properties
of $^3$H and $^4$He with the NNLO* potential
and find good agreement with the data and with 
predictions based upon the 
standard high--precision potentials. We find an improved description of the 
$^3$H and $^4$He binding energies.
\end{abstract}

\vfill

\pagebreak


\section{Introduction}
\def\theequation{\arabic{section}.\arabic{equation}}
\setcounter{equation}{0}

Nuclear forces are derived in the chiral effective field theory approach 
in terms of an expansion in powers of $Q/\Lambda_\chi$, where $Q$ corresponds to
a generic external momentum of nucleons and $\Lambda_\chi$ represents 
the typical hadronic scale (scale of chiral symmetry breaking) 
of the order of 1 GeV. 
That ratio is less than one if one considers processes with sufficiently low 
external momenta of the nucleons. 
In order to exclude contributions of high--momentum components in intermediate states, the 
nucleon--nucleon (NN) 
potential is multiplied by a regulator, which suppresses  momenta 
larger than a certain  cut--off $\Lambda$ \cite{Lepage}. The latter has to be chosen below the scale 
$\Lambda_\chi$.\footnote{In some cases it turns out to be possible to perform standard
renormalization of the theory by taking the cut--off $\Lambda$ to infinity \cite{LimC,EMprep}.}
The cut--off $\Lambda$ should also not be taken too small in order not to suppress the relevant
physics.
The various coupling constants depend on the cut--off $\Lambda$ in a way to compensate 
the changes in the low--energy observables induced by varying $\Lambda$. 
The remaining cut--off dependence of  the 
observables can be removed by adding higher order terms to the effective potential \cite{Lepage}. 
Assuming naturalness for the various renormalized coupling constants in the
underlying Lagrangian one can expect that contributions to the NN forces
corresponding to higher powers $\nu$ of the chiral expansion  
will decrease. This sort of  nuclear interactions based on
the most general chiral invariant 
effective Lagrangian formed out of pion and nucleon fields has been
first proposed in \cite{Weinb1} and formulated in detail in \cite{bira1}. We followed a
similar path, however  extracting the nuclear forces from the Lagrangian in
a different way. We refer to \cite{egm1} where two-- and three-- nucleon
potentials have been derived using the method of unitary
transformation. That method leads to energy independent and hermitean
nuclear forces which are better suited for applications to systems with $A > 2$
than energy-dependent forces derived in old fashioned
time--ordered perturbation theory like in \cite{bira1}. In \cite{ourPRL} we applied the forces
at next--to--leading order (NLO), corresponding to (the counting index) $\nu=2$, to the
3N and 4N systems. At this order NN phase shifts can be described only at rather
low energies and only modestly. Nevertheless 3N and 4N
binding energies were found to be within the same range as the ones found with high
precision modern NN forces and also nd elastic and break-up observables
at very low energies
are similar to predictions generated by conventional forces. At that
order the  experimental nucleon analyzing power $A_y$ is fairly well
reproduced, which
for conventional NN forces poses a serious puzzle \cite{GloeckleREP}. This result, however, has to
be considered as an intermediate step, corresponding just to NLO,
where the $^3P_j$ NN phase shifts could not be reproduced with sufficient
accuracy. It is now of strong interest to explore the chiral forces in 3N
and 4N systems at next--to--next--to--leading order (NNLO) corresponding to  $\nu=3$
where the NN
phase shifts are better reproduced. For the convenience of the
reader we review briefly the NN forces in LO ($\nu=0$), NLO and NNLO in section
2. 

It has been already pointed out in \cite{egm2}  that the strong central attraction
caused by the numerically large values of the LEC's $c_1$, $c_3$, and $c_4$ as determined in 
a $Q^3$ analysis of $\pi$N 
scattering leads to spurious deeply bound states in various two--nucleon 
angular momentum states. Though this has no observable
consequences in the NN system within the realm of validity of the theory
it is technically somewhat disturbing in treating 3N and 4N systems. Also
ignoring 3N forces, which occur at NNLO the first time, and
exploring only the NNLO NN forces leads to strong deviations from 3N
data as we will show. It has to be expected that this will be remedied by including the
NNLO 3N forces, which necessarily have to be taken into account at that
order. 
Various consequences of the large values of the $c_i$'s 
as well as the current situation in relation to the  
determination of the $c_i$'s from other processes (such as $\pi$N scattering)
are discussed in section 3.
Motivated by the findings of the boson--exchange (BE) models of the 
nucleon-nucleon interaction, we constructed the NNLO* potential by 
removing the $\Delta$ content from the LEC's  $c_3$ and $c_4$ and 
refitting the contact interactions. 
The new values of the $c_i$'s resulting from subtracting the $\Delta$ contributions 
lead to the NNLO* potential
which is free of spurious NN
bound states for the cut--off range considered. The resulting NN phase shifts as
shown in section 3 are significantly improved as compared to the NLO result. 
We also discuss in this section various deuteron properties.
It should be mentioned that all these conclusions are based on the
type of regulator we employ in the Lippmann-Schwinger equation. It
cannot be excluded at present that a regulator can be constructed that
allows for using the large $c_i$ without leading to deep virtual bound
states. However, if such regularization exists, it has to look very
different than the commonly employed regulator functions.

We then switch to the 3N and 4N systems and briefly demonstrate in section 4
the predictions corresponding to the NNLO potential. As already stated before,
neglecting the 3N forces leads to strong deviations from the data. 

The central results of our paper, namely the application of the NNLO* potential
to predict 3N and 4N observables, are presented in section 5.
All these results have to be
supplemented in the future 
by the inclusion of the three types of
topologically different  3N forces which occur at NNLO. This additional
extensive
investigation is left to a forthcoming  paper. We summarize briefly in
section 6.

\section{Few--Nucleon Forces in Chiral Effective Field Theory}
\setcounter{equation}{0}

Starting from the most general chiral invariant effective Hamiltonian
density for pions and nucleons one can derive nuclear forces by
eliminating the pions through a method of unitary transformation \cite{egm1}. Since
this transformation acts on the field theoretical Hamiltonian, it leads
to an energy--independent effective Hamiltonian in the pure nucleonic
space. The condition for decoupling the purely nucleonic Fock space states
from the ones with pions, a nonlinear decoupling equation, can
be linearized by introducing a series of orthonormal subspaces with
different number of pions leading to an infinite set of coupled
equations determining the unitary operator. Those equations can be solved
recursively. Thereby the basic organization principle is a counting scheme in
powers of momenta and number of pions. We refer to \cite{egm1} for the detailed steps.
Notice also that the relativistic $1/m$ corrections are assumed
to be suppressed compared to the $1/\Lambda_\chi$ ones, see \cite{Weinb1}.
Further, we will consider only the isospin invariant case in this section.
Isospin violating effects can be treated along the lines presented in refs.\cite{Bira_phd},\cite{WME}.
The resulting nucleonic potentials are ordered by the power
\beq
\label{powc}
\nu = -4 + E_n + 2 L + \sum_i V_i \Delta_i~,
\eeq
where $E_n$, $L$ and $V_i$ are the numbers of external nucleon lines, loops and vertices of
type $i$, respectively. Further, the quantity $\Delta_i$, which defines the dimension of a vertex of
type $i$,  is given by
\beq
\label{deltai}
\Delta_i = d_i + \frac{1}{2} n_i - 2~,
\eeq
with $d_i$ the number of derivatives or $M_\pi$ insertions and $n_i$ the number of 
nucleon lines at the vertex $i$.  The inequality $\Delta_i \geq 0$ holds true 
as a consequence of chiral invariance.
This leads to $\nu \geq 0$ for processes with two and more nucleons. 
One also recognizes that the diagrams with loops are suppressed and that $(n+1)$--nucleon 
forces appear at higher orders than $n$--nucleon forces.

Let us now consider first several orders of the NN force.
\begin{figure}[htb]
\vspace{1.2cm}
\centerline{
\psfig{file=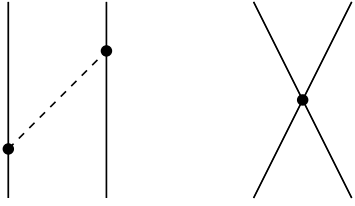,width=1.858in}}
\vspace{0.3cm}
\centerline{\parbox{14cm}{
\caption[fig1]{\label{fig1}  
Leading order (LO) contributions to the NN potential: one--pion exchange and contact diagrams.
Graphs which result from the interchange of the 
two nucleon lines are not shown. Solid and dashed lines are nucleons and pions, respectively. The heavy
dots denote the vertices with $\Delta_i =0$.}}}
\vspace{0.7cm}
\end{figure}
At leading order $\nu=0$ (LO) only tree diagrams with vertices
of $\Delta_i = 0$ ($\pi$NN vertex with one derivative and two independent four--nucleon contact 
interactions without derivatives) are allowed, see eq.~(\ref{powc}). Consequently, the LO 
chiral potential is given by the well established one--pion exchange (OPE) and contact forces with the 
low energy constants (LEC) $C_S$ and $C_T$, as shown
in Fig.~\ref{fig1}:
\beqa
\label{vlo}
V^{(0)}_{\rm cont} &=&  C_S  + C_T \; \sigma_1 \cdot \sigma_2 \;, \\
V^{(0)}_{\rm OPEP} &=& -\biggl(\frac{g_A}{2f_\pi}\biggr)^2 \, \fet{\tau}_1 \cdot
\fet{\tau}_2 \, \frac{\vec{\sigma}_1 \cdot\vec{q}\,\vec{\sigma}_2\cdot\vec{q}}
{q^2 + M_\pi^2}~. \nonumber
\eeqa
Here $\vec p$ and $\vec{p} \, '$ are the initial and final momenta of the nucleons in the CM frame
and $\vec{q} = \vec{p}\, '- \vec{p}$. Further, $M_\pi$, $g_A$, and $f_\pi$ are the pion mass, the axial 
pion--nucleon coupling constant and the pion decay constant, respectively.

\begin{figure}[htb]
\vspace{1.2cm}
\centerline{
\psfig{file=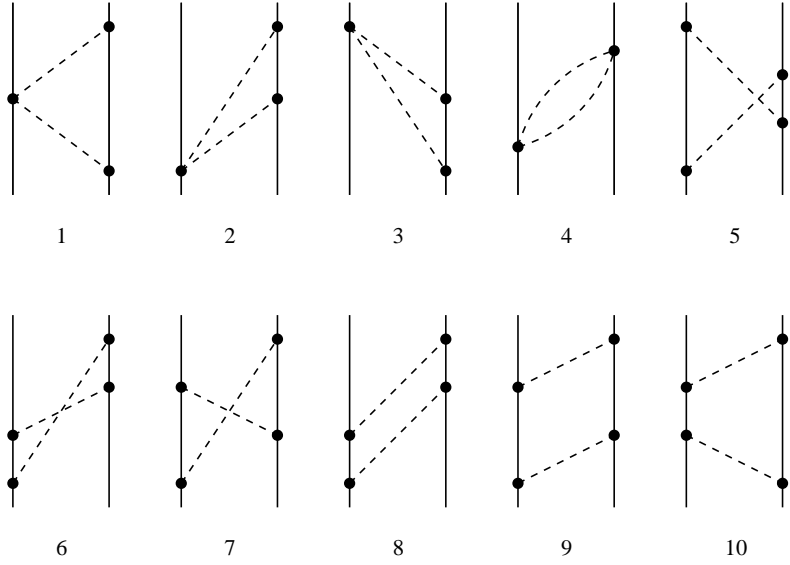,width=4.1125in}}
\vspace{0.3cm}
\centerline{\parbox{14cm}{
\caption[fig2]{\label{fig2}  First corrections at NLO to the NN potential in the projection
formalism: two--pion exchange diagrams.
For notations see fig.~\ref{fig1}.}}}
\vspace{0.7cm}
\end{figure}

\begin{figure}[htb]
\vspace{1.2cm}
\centerline{
\psfig{file=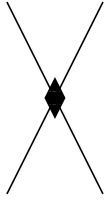,width=0.533in}}
\vspace{0.3cm}
\centerline{\parbox{14cm}{
\caption[fig3]{\label{fig3}  
First corrections to the NN potential: contact diagram at 
next--to--leading order (NLO). The filled diamond denotes seven vertices 
of $\Delta_i=2$ (with two derivatives). For remaining notations see fig.~\ref{fig1}.}}}
\vspace{0.7cm}
\end{figure}

At next--to--leading  order (NLO)
or $\nu=2$ there are TPE diagrams with the leading $\pi$NN vertices with $\Delta_i=0$ according
to Fig.~\ref{fig2} and seven contact forces with vertices of $\Delta_i=2$  containing two
derivatives\footnote{The 
contact interactions with one insertion of $M_\pi^2$ are formally indistinguishable 
from the  four--nucleon operators without derivatives  and lead to renormalization of 
the constants $C_S$, $C_T$. We will not consider such operators explicitly.}, see Fig.~\ref{fig3}.  
It should be emphasized at this stage, that the expression (\ref{powc}) only allows to estimate the 
order of the corresponding process. It is, however, not possible to read off the precise structure of the 
operators (i.e.~the corresponding energy denominators and overall factors) related to a particular diagram. 
This is because the presented figures refer to  diagrams within the method of unitary transformation and
not to ordinary graphs in the old--fashioned perturbation theory. The precise operator form of the NLO and 
NNLO contributions to the 2N and 3N potentials can be found in reference \cite{egm1}. Note also that the graphs 
9 and 10 in fig.~\ref{fig2} are not reducible ones in the sense, that no energy denominators related to purely
nucleonic intermediate states appear in the corresponding expressions; see \cite{egm1} for more details.  

In addition, there are nucleon self--energy contributions and vertex
corrections \cite{egm1},  which renormalize the one--pion exchange and contact
forces, which we do not show explicitly here. 
The TPE terms shown in Fig.~\ref{fig2} lead to polynomial parts with, in general, infinite
coefficients, which renormalize various contact interactions, and to 
finite non--polynomial ones, which are finite and 
independent of the regularization scheme used. 
The resulting potential
reads:
\beqa
\label{vnlo}
V^{(2)}_{\rm cont} &=& C_1 \, \vec{q}\,^2 + C_2 \, \vec{k}^2 +
( C_3 \, \vec{q}\,^2 + C_4 \, \vec{k}^2 ) ( \vec{\sigma}_1 \cdot \vec{\sigma}_2)
+ iC_5\, \frac{1}{2} \, ( \vec{\sigma}_1 + \vec{\sigma}_2) \cdot ( \vec{q} \times
\vec{k})\nn
&& {} +  C_6 \, (\vec{q}\cdot \vec{\sigma}_1 )(\vec{q}\cdot \vec{\sigma}_2 ) 
+ C_7 \, (\vec{k}\cdot \vec{\sigma}_1 )(\vec{k}\cdot \vec{\sigma}_2 )\;,\\
V^{(2)}_{\rm TPEP} 
&=& - \frac{ \fet{\tau}_1 \cdot \fet{\tau}_2 }{384 \pi^2 f_\pi^4}\,
L(q) \, \biggl\{4M_\pi^2 (5g_A^4 - 4g_A^2 -1) + q^2(23g_A^4 - 10g_A^2 -1)
+ \frac{48 g_A^4 M_\pi^4}{4 M_\pi^2 + q^2} \biggr\}\nn
&& {} - \frac{3 g_A^4}{64 \pi^2 f_\pi^4} \,L(q)  \, \biggl\{
\vec{\sigma}_1 \cdot\vec{q}\,\vec{\sigma}_2\cdot\vec{q} - q^2 \, 
\vec{\sigma}_1 \cdot\vec{\sigma}_2 \biggr\} ~, \nonumber
\eeqa
where
\beq
\label{res1}
L(q) = \frac{1}{q}\sqrt{4 M_\pi^2 + q^2}\, 
\ln\frac{\sqrt{4 M_\pi^2 + q^2}+q}{2M_\pi}~,
\eeq
and 
$\vec{k} = 1/2 (\vec{p}\, '+ \vec{p})$.
There are seven LEC's $C_1$ to $C_7$ related to contact interactions with two derivatives, see fig.~\ref{fig3}.

\begin{figure}[htb]
\vspace{1.2cm}
\centerline{
\psfig{file=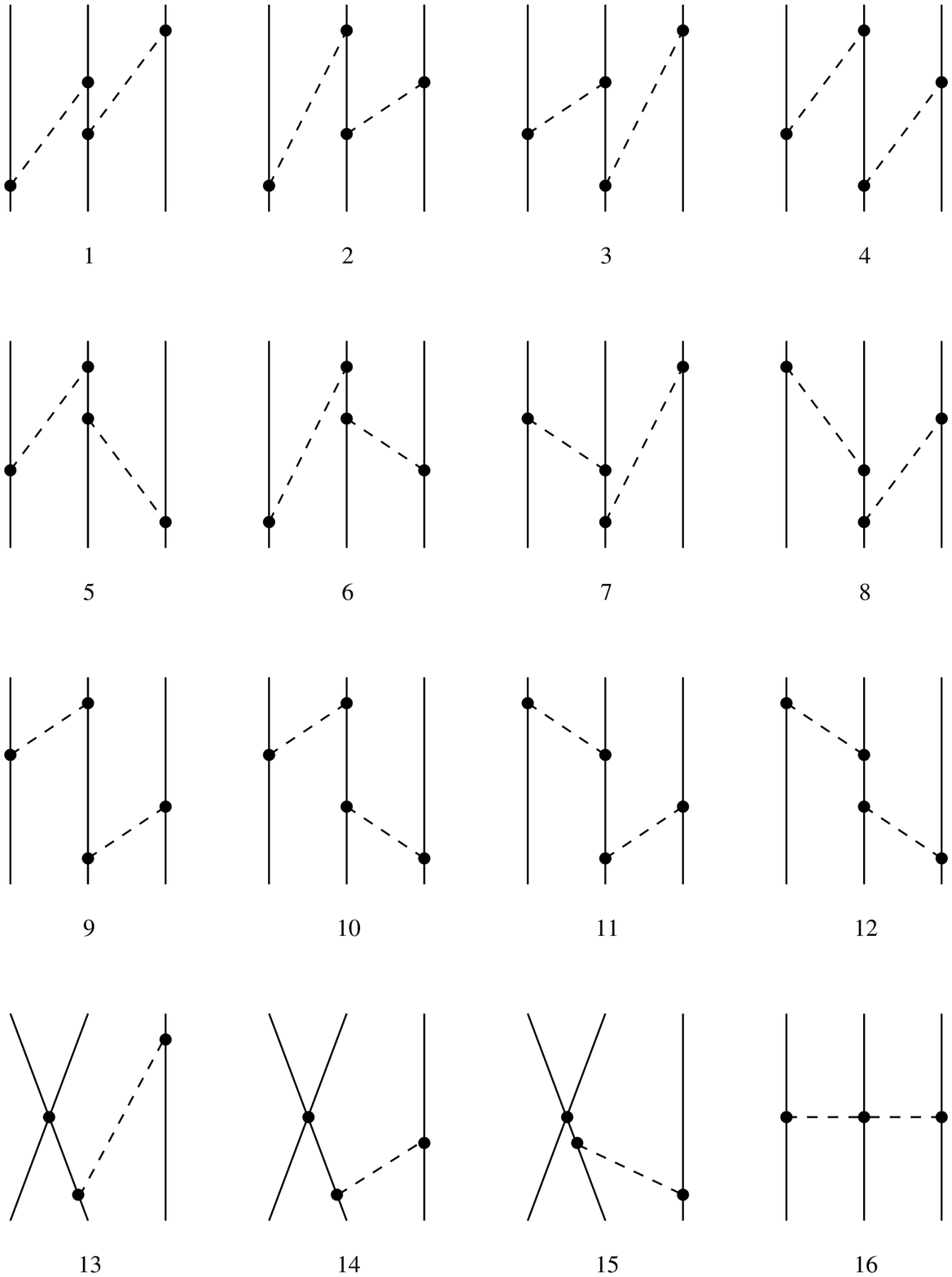,width=4.652in}}
\vspace{0.3cm}
\centerline{\parbox{14cm}{
\caption[fig4]{\label{fig4} Leading contributions to the three--nucleon potential at NLO, which cancel: 
two--pion and one--pion exchange 
diagrams with the NN contact interaction. Graphs which result from the 
interchange of the nucleon lines and/or from the application of time reversal operation are not shown. 
In the case of diagram 16, one should
sum over all possible time orderings. For remaining notations see fig.~\ref{fig1}.}}}
\vspace{0.7cm}
\end{figure}

At that order NLO 3N forces of the
topologies shown in Fig.~\ref{fig4} cancel.
Note that this cancellation is of different
type than the one found in time--ordered perturbation theory \cite{Yang}, \cite{biraX}.
To be more precise, in that order the contribution of the ``irreducible'' two--pion (one--pion) exchange diagrams 
1--8 (13) cancels against the ``reducible'' two--pion (one--pion) exchange graphs 9-12 (14, 15). 
The last graph 16 in this figure is proportional to the kinetic energy of the nucleons and contributes therefore 
only at higher orders \cite{Weinb1}. 

\begin{figure}[htb]
\vspace{1.2cm}
\centerline{
\psfig{file=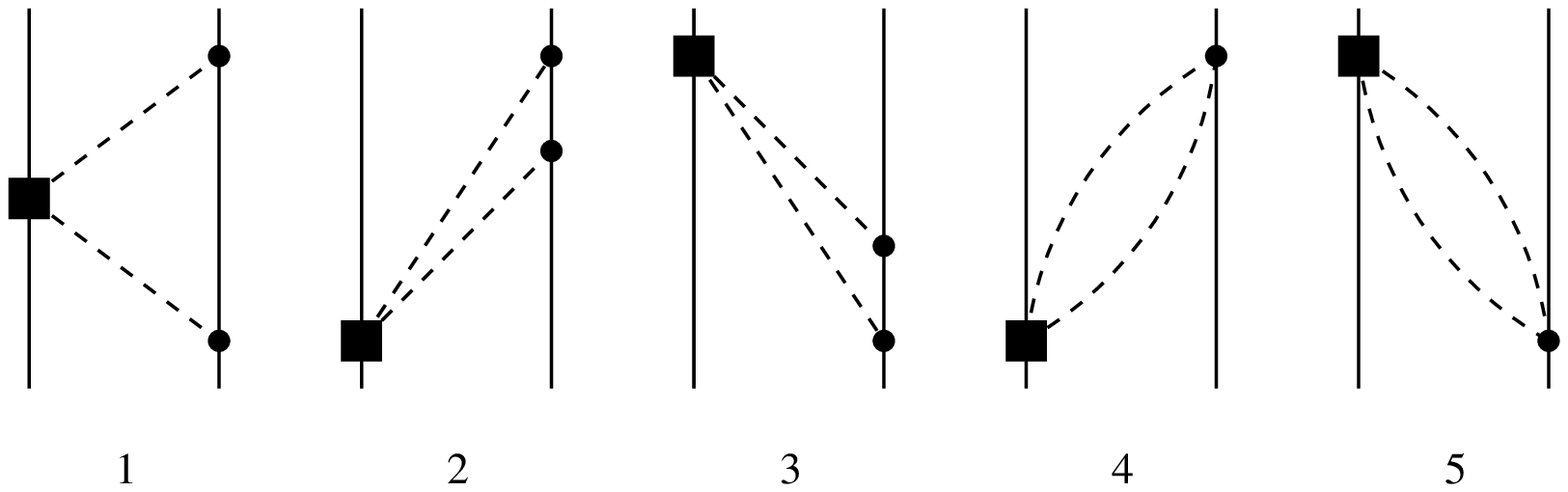,width=4.229in}}
\vspace{0.3cm}
\centerline{\parbox{14cm}{
\caption[fig5]{\label{fig5}  Next--to--next--to--leading order (NNLO) 
corrections to the NN potential. 
The filled squares denote the vertices with $\Delta=1$. 
For remaining notations see figs.~\ref{fig1}, \ref{fig3}.}}}
\vspace{0.7cm}
\end{figure}

\begin{figure}[htb]
\vspace{1.2cm}
\centerline{
\psfig{file=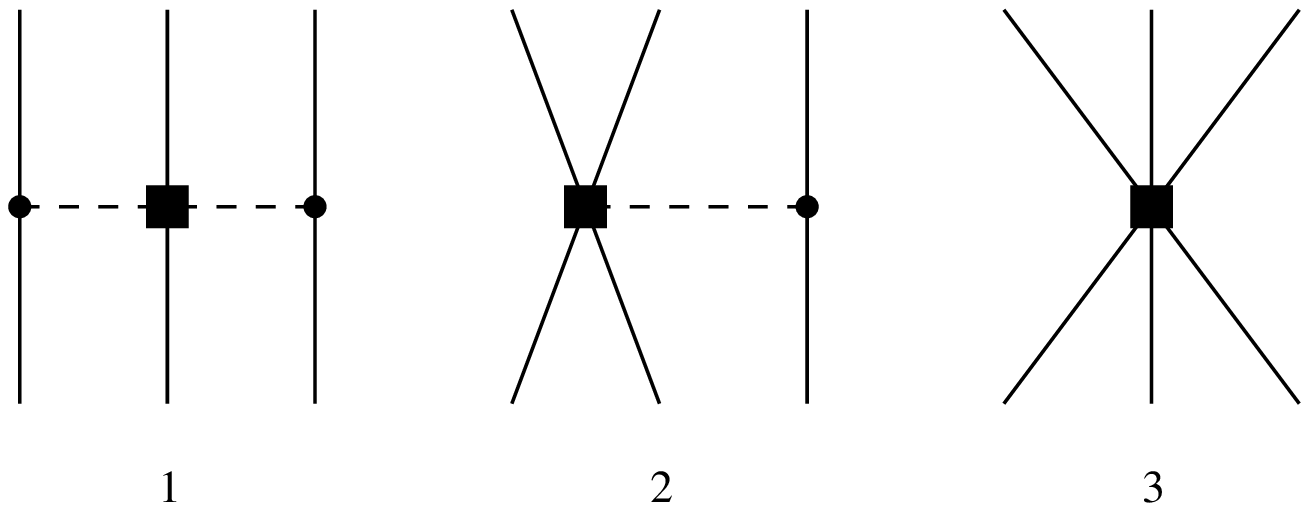,width=3.352in}}
\vspace{0.3cm}
\centerline{\parbox{14cm}{
\caption[fig6]{\label{fig6} Three--nucleon force: TPE, OPE and contact interaction.
In the cases of diagrams 1 and 2, all possible time orderings should be taken into account. 
For notations see figs.~\ref{fig1} and \ref{fig5}.}}}
\vspace{0.7cm}
\end{figure}

At NNLO ($\nu=3$) there occur new $\pi \pi$NN vertices with $\Delta_i=1$, which contain
either two derivatives or one  $M_\pi^2$ insertion and are parametrized by three
constants,  denoted in the commonly used notation by  $c_1$, $c_3$, and $c_4$ (the $c_2$--term
does not contribute at this order) \cite{BKMrev}.
They enter into the TPE NN force as shown in Fig.~\ref{fig5} as well as into
vertex correction diagrams (not shown), which
renormalize the OPE, and also into the TPE 3N force shown
in Fig.~\ref{fig6}. 
The explicit expression for the two--pion exchange NN force at NNLO is\footnote{Note that we
included here the $1/m$ corrections, which are formally of the higher order.}
\beqa
\label{vnnlo}
V^{(3)}_{\rm TPEP} &=& -\frac{3g_A^2}{16\pi f_\pi^4} \biggl\{ -\frac{g_A^2
  M_\pi^5}{16 m (4M_\pi^2+q^2)} + \biggl(2M_\pi^2(2c_1 -c_3) -q^2 \,
\bigl( c_3 + \frac{3g_A^2}{16m} \bigr) \biggr) (2M_\pi^2+q^2) A(q) \biggr\}
\nn
&-& \frac{g_A^2}{128\pi m f_\pi^4} (\fet{ \tau}_1 \cdot \fet{ \tau}_2 ) \,
\biggl\{ -\frac{3g_A^2  M_\pi^5}{4M_\pi^2+q^2} + \bigl( 4M_\pi^2 +
2q^2 -g_A^2 (4M_\pi^2 + 3q^2) \bigr)  (2M_\pi^2+q^2) A(q) \biggr\}
\nn
&+&  \frac{9g_A^4}{512\pi m f_\pi^4} \biggl( (\vec \sigma_1 \cdot \vec
q\,)(\vec \sigma_2 \cdot \vec q\,) -q^2 (\vec \sigma_1 \cdot\vec \sigma_2
)\biggr) \, (2M_\pi^2+q^2) A(q)  \nn
&-& \frac{g_A^2}{32\pi f_\pi^4} (\fet{ \tau}_1 \cdot \fet{ \tau}_2 ) \,
\biggl( (\vec \sigma_1 \cdot \vec q\,)(\vec \sigma_2 \cdot \vec q\,) 
-q^2 (\vec \sigma_1 \cdot\vec \sigma_2 )\biggr) \nn
&& \qquad   \qquad \qquad \qquad   \qquad \qquad \times
\biggl\{ \bigl( c_4 + \frac{1}{4m} \bigr) (4M_\pi^2 + q^2) 
-\frac{g_A^2}{8m} (10M_\pi^2 + 3q^2) \biggr\} \, A(q) \nn
&-& \frac{3g_A^4}{64\pi m f_\pi^4} \, i \, (\vec \sigma_1 +  \vec
\sigma_2 ) \cdot (\vec{p}~' \times \vec{p} ) \, (2M_\pi^2+q^2) A(q) \nn
&-& \frac{g_A^2(1-g_A^2)}{64\pi m f_\pi^4} (\fet{ \tau}_1 \cdot \fet{ \tau}_2 )
\, i \, (\vec \sigma_1 +  \vec \sigma_2 ) \cdot (\vec{p}~' \times
\vec{p} ) \, (4M_\pi^2+q^2) A(q) \;, \nonumber
\eeqa
where 
\beq
\label{res2}
A(q) = \frac{1}{2q} \arctan \frac{q}{2M_\pi}~. 
\eeq
Altogether there are 9 LEC's at NNLO (and at NLO) 
related to various contact interactions, which have to be
fitted by adjusting the NN force to the NN data. 
The  LEC's $c_{1,3,4}$ which first appear at NNLO occur also in
$\pi$N scattering and that information should be consistently taken
into account.

The 3N force at NNLO consists of three different topologies as shown in fig.~\ref{fig6}. 
Besides the TPE there is a pion
exchange between a NN contact force and the third nucleon and a pure 3N
contact force. 
In both cases new vertices of $\Delta_i=1$ with unknown constants enter.
The precise structure of the chiral 3NF will be discussed in a forthcoming paper.

Chiral forces are only valid in a low--momentum region. We enforce this by
modifying the above given NN force expressions as
\beq\label{Vreg}
V( \vec{p}~' ,\vec{p}\,) \to f_R ( \vec{p}~' \,) \, V( \vec{p}~',\vec{p}\,) \,
f_R (\vec{p}\,)~,
\eeq
where $f_R ( \vec{p}\,)$ is a regulator function. 
In what follows, we work with the following regulator function:
\beq\label{reg1}
 f_R^{\rm expon} ( \vec{p}\,) = \exp(-p^{4} / \Lambda^{4})~.
\eeq
The power four in the exponent guarantees that  the $Q^0$--, $Q^2$-- and $Q^3$--terms in the 
potential are not affected by the regularization procedure.
As already pointed out before, the dependence of the low--energy observables on the value 
of the cut--off $\Lambda$ should get weaker with increasing
the order $\nu$.

\section{Two Nucleons at Next--to--Next--to--Leading Order}

We now turn to the analysis of the 2N system at NNLO.
Let us first specify the parameters entering the NN potential. 
The largest uncertainty is related to contact interactions between nucleons. They are not restricted
by chiral symmetry, but only by the general principles of locality, invariance under 
Lorentz transformations, parity, time--reversal invariance and
hermiticity. At NLO and NNLO one has to take into 
account nine independent
contact operators contributing to the effective potential: two operators without derivatives 
($V^{(0)}_{\rm cont}$ in eq.~(\ref{vlo})) and seven
with two derivatives of nucleon fields ($V^{(2)}_{\rm cont}$ in eq.~(\ref{vnlo})). The corresponding LECs 
are fixed by a fit to S-- and P--wave phase shifts and to $\epsilon_1$ at low
energies. The OPE ($V^{(0)}_{\rm OPEP}$ in eq.~(\ref{vlo})) as well as the leading chiral TPE at NLO 
($V^{(2)}_{\rm TPEP}$ in eq.~(\ref{vnlo})) are parameter--free. 

As already stressed before, the subleading TPE at NNLO, $V^{(3)}_{\rm TPEP}$ in eq.~(\ref{vnnlo}),  
depends on the LECs $c_{1,3,4}$, which correspond to 
$\pi \pi$NN vertices of dimension $\Delta_i=1$. Precise numerical values for these constants are 
crucial for various properties of the effective NN interaction as will be discussed below. 
Clearly, the subleading $\pi \pi$NN vertices 
represent an important link between NN scattering and other processes, such as $\pi$N scattering. Therefore, ideally, 
one would like to take their values from the analysis of the $\pi$N system, 
as it was done in \cite{egm2}.
We will now briefly overview the current situation concerning the
determination of the $c_i$'s from the $\pi$N system.
Several calculations
for $\pi$N scattering have been performed and published. From the $Q^2$ analysis \cite{BKM95} one gets:
$c_1 = -0.64\,,  c_3 = -3.90\,, c_4=2.25\;$.
Here  all values are given in GeV$^{-1}$.
{}From different $Q^3$ calculations \cite{BKM95}, \cite{BKM97}, \cite{Moi98}, \cite{FMS99}, \cite{Paul} one obtains
the following bands for the $c_i$'s:
\begin{equation} 
\label{q3}
c_1 = -0.81 \ldots -1.53 \,, \quad \quad c_3 = -4.70 \ldots -6.19\,, \quad \quad c_4 = 3.25 \ldots 4.12\,.
\end{equation}
These bands are also consistent with expectations from resonance saturation, see \cite{BKM97}.
Recently, the results from a $Q^4$ analysis have become available \cite{FM00}.
At this order the S--matrix is sensitive to 14 LECs (including $c_{1,3,4}$), which 
have been fixed from a fit to $\pi$N phase shifts. 
At this order the dimension two LECs acquire a quark mass
renormalization. The corresponding shifts are proportional to
$M_\pi^2$. It turns out that different
phase shift analyses (PSA) from refs.~\cite{Koch86}, \cite{Mat97} and \cite{SAID} 
lead to sizable variations in the actual values of the LECs. 
A typical fit  
based on the  phases of ref.~ \cite{Mat97} leads to:
\begin{equation}
\label{mats}
\tilde{c}_1 = -0.27 \pm 0.01\,, \quad \quad \tilde{c}_3 = -1.44 \pm 0.03\,, \quad \quad \tilde{c}_4=3.53
\pm 0.08 \;,
\end{equation} 
where $\tilde{c}_i$ denote the renormalized $c_i$'s.
However, using the older Karlsruhe or the VPI phases as input, one
finds sizable variations in the $\tilde{c}_i$.  Alternatively, one can
also keep the $c_i$ at their third order values and fit the fourth
order corrections separately, see \cite{FM00}. Due to the  
uncertainties in the isoscalar amplitudes, these constants are not very
well determined. The fits could, in principle, be improved in the future by
including the scattering lengths determined from pionic hydrogen/deuterium.
To complete the discussion on determination of the $c_i$'s from the $\pi$N system
we would like to stress, that
numbers consistent with the bands given
in eq.(\ref{q3}) have been obtained in \cite{BL} using IR regularized
baryon chiral perturbation theory at order $Q^3$ and dispersion relations. 

Rentmeester et al.~\cite{Rent} 
tried to fix the values of the $c_i$'s from  an analysis of the $pp$ data,
which are of a much better quality than the $\pi$N data.
In this approach the long--range part of the NN force was taken as the sum 
of the OPE and the chiral TPE (including the NNLO contribution). The  
NN interaction at short distances below some boundary value was parametrized by some artificial 
energy dependent representation.
The global fit to the data allowed to pin down the values of the $c_i$'s (and, of course,
also of the parameters related to the short--range part of the NN force). It turned out that it is  
not possible to fix all three $c_i$'s in 
this process because of the strong correlation between these LECs. 
For that reason the constant $c_1$ was fixed at the value $c_1 =-0.76\,$GeV$^{-1}$
(to obtain a small pion-nucleon $\sigma$-term of about 40~MeV)
and the LECs $c_{3,4}$ were treated as free parameters.
The values of the $c_{3,4}$: $c_3 = -5.08$ GeV$^{-1}$, $c_4=4.70$ GeV$^{-1}$ 
determined from the global fit to 
the $pp$ data are compatible with the 
$Q^3$ calculation from the $\pi$N system, see eq.~(\ref{q3}). Note, however, that this method
is not directly based on a systematic chiral power counting.   

Having overviewed the current status of the determination of the $c_i$'s from various processes,
we are now in the position to discuss the corresponding implications for the NN system.
First of all, it turns out that the  numerical values of
the $c_i$'s  
are quite large. Indeed, from a dimensional analysis one would expect, for example, the constant $c_3$ 
to scale like:
\begin{equation}
\label{scaling}
c_3 \sim \frac{{\ell}}{ 2\Lambda_\chi} \,,
\end{equation}
where ${\ell}$ is some number of order one. Taking the value $c_3 = - 4.70$ from ref.~\cite{Paul}
and $\Lambda_\chi = M_\rho = 770$ MeV we end up with ${\ell} \sim - 7.5$. Such a large value 
can be partially explained by the fact that the $c_{3,4}$ 
are to a large extent saturated by the $\Delta$--excitation. This implies that a new and smaller 
scale, namely $m_\Delta - m \sim 293$ MeV, enters the values of these constants, see \cite{BKM97}.

What are the consequences  of the large numerical values of the $c_i$'s for NN scattering?
The main problem is that the large numerical values of the $c_i$'s might lead to a slow 
convergence of the low--momentum expansion. To get a feeling of the possible problems one can
compare, for instance, the  low--momentum matrix elements of, say, the central parts of the TPE at 
NLO and NNLO. Taking the values of the $c_i$'s from the $Q^3$--analysis of the $\pi$N system from ref.~\cite{Paul}
\beq
\label{ci_central}
c_1=-0.81\;  \mbox{GeV}^{-1}\, , \quad \quad   
c_3=-4.70\;  \mbox{GeV}^{-1}\, , \quad \quad   
c_4=3.40\;  \mbox{GeV}^{-1}\, ,   
\eeq
as we did in \cite{egm2} one gets 
from eqs.~(\ref{vnlo}), (\ref{vnnlo}):
\beqa
V_{\rm TPE}^{\rm cent, \;(2)} (q)\bigg|_{q=0} &=&  (\fet \tau_1 \cdot \fet \tau_2)\, \frac{M_\pi^2}{(4 \pi f_\pi)^2 f_\pi^2}
\; \frac{(1 + 4 g_A^2 - 8 g_A^4)}{6} \sim ( \fet \tau_1 \cdot \fet \tau_2 ) (-3.4) \;\mbox{GeV}^{-2}\,, \\
V_{\rm TPE}^{\rm cent, \;(3)} (q)\bigg|_{q=0} &=& \frac{M_\pi^2}{(4 \pi f_\pi)^2 f_\pi^2} \; (-3 g_A^2 \pi) \; (2 c_1 - c_3) 
\; M_\pi  \sim -10.3  \;\mbox{GeV}^{-2} \;. \nonumber
\eeqa
Here we neglected all $1/m$--corrections.
While the order of the matrix element of the potential at NLO agrees with the one expected from dimensional
analysis, $V_{\rm TPE}^{\rm cent, \;(2)} \sim ( \fet \tau_1 \cdot \fet \tau_2)\; 
\ell_1 M_\pi^2 / (\Lambda_\chi^2 f_\pi^2)$ with $\ell_1 \sim -0.9$, the NNLO matrix element appears to be 
larger than expected: $V_{\rm TPE}^{\rm cent, \;(3)} \sim  \ell_2 M_\pi^3 / (\Lambda_\chi^3 f_\pi^2)$ 
where $\ell_2 \sim -14.3$. Such a deviation from the natural value for $\ell_2$ of order one
does, however,  not yet
necessarily mean a failure of the perturbative expansion, since the potential itself is not an observable 
quantity. To draw a precise conclusion about the convergence properties of the low--momentum expansion 
one should look at the phase shifts, which can be measured directly. Further, up to now we only compared the non--polynomial 
contributions to the potential and omitted all contact terms.\footnote{Note that the contact interactions
are needed to renormalize the TPE contribution and thus cannot be omitted for conceptual reasons.} 
Large numerical values of the low--momentum 
matrix elements of the $V_{\rm TPE}$ at NNLO could, in principle, be compensated by the corresponding contact terms.
\begin{figure}[hbt]
\begin{center}
\psfrag{1D2}{\raisebox{-0.2cm}{\hskip -0.0 true cm  $^1D_2$}}
\psfrag{3D2}{\raisebox{-0.2cm}{\hskip -0.0 true cm  $^3D_2$}}
\psfrag{xxx}{\raisebox{0.0cm}{\hskip -0.7 true cm  $E_{\rm lab}$ [GeV]}}
\psfrag{yyy}{\raisebox{0.0cm}{\hskip -0.3 true cm  $\delta$ [deg]}}
\parbox{7cm}{
\centerline{\hskip 0.3 true cm \psfig{file=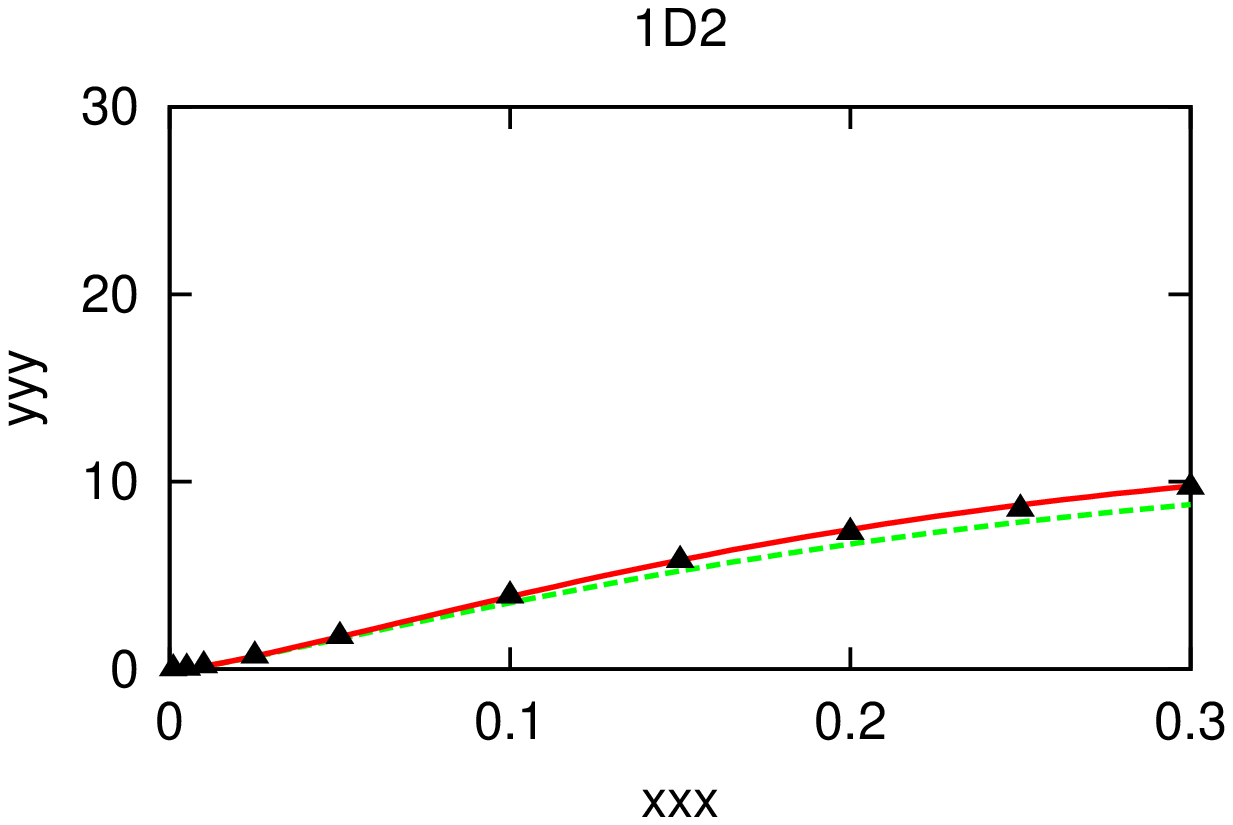,width=7.5cm,height=4.7cm}}}
\hskip 1 true cm 
\parbox{7cm}{
\centerline{\hskip -0.0 true cm \psfig{file=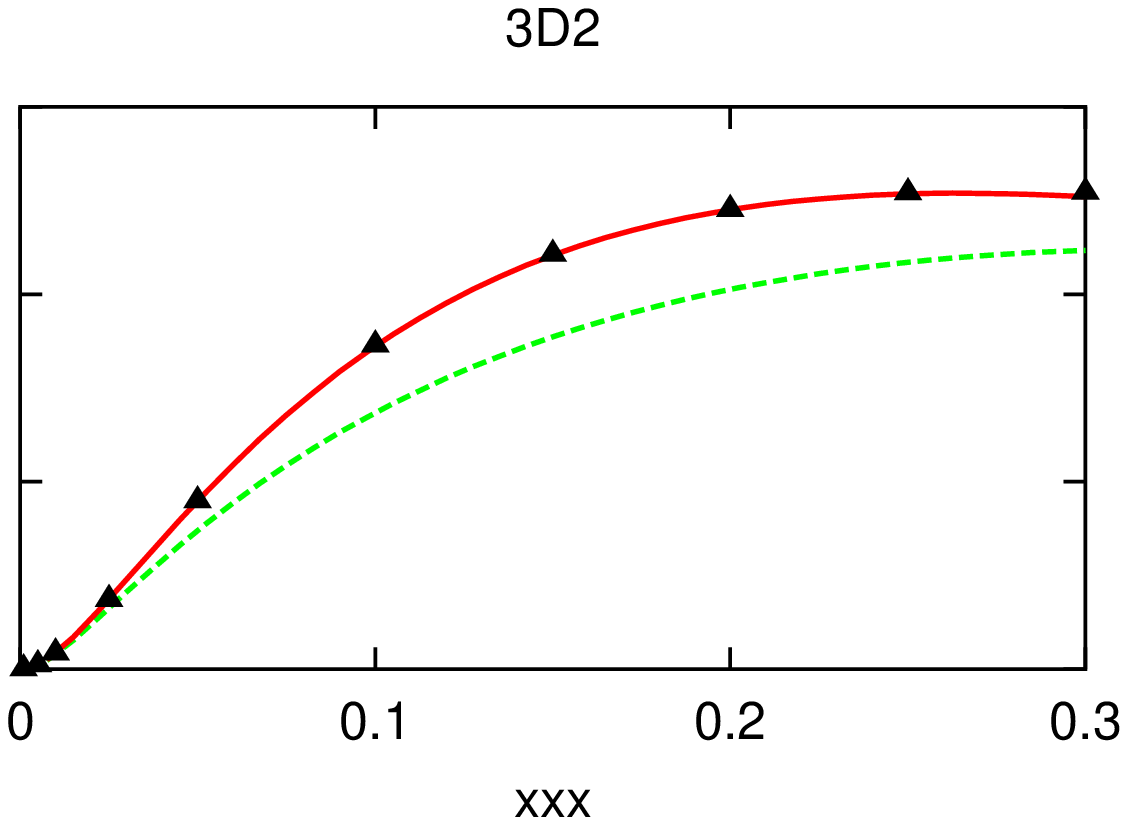,width=6.9cm,height=4.7cm}}}
\centerline{\parbox{14cm}{
\caption{\label{fig:dw_cdb}$^1D_2$ and $^3D_2$ phase shifts calculated with the CD-Bonn potential.
The dashed lines show the Born approximation, whereas the solid lines correspond to the full  
solution of the Lippmann--Schwinger equation. The filled triangles are Nijmegen PSA results \cite{NPSA}}}}
\end{center}
\end{figure}
However, such a compensation at NNLO is only possible 
for S-- and P--waves as well as for $\epsilon_1$ since the contact terms do not contribute to 
D-- and higher partial waves at this order.
The D-- and F--waves may therefore serve as a sensitive test of the chiral TPE exchange.\footnote{This 
has been suggested by Kaiser et al. in \cite{KBW}, \cite{KGW}.}
The conventional scenario of nuclear forces represented by existing OBE models and various phenomenological
potentials suggests that the  D-- and higher partial wave NN interactions are weak enough to be treated perturbatively. This is demonstrated in fig.~\ref{fig:dw_cdb} on the example of the CD-Bonn potential. 
Although this observation is confirmed by the smallness of the corresponding phase shifts, such a scenario, 
strictly speaking,
does not necessarily need to be realized.
\begin{figure}[hbt]
\begin{center}
\psfrag{1D2}{\raisebox{-0.2cm}{\hskip -0.0 true cm  $^1D_2$}}
\psfrag{3D2}{\raisebox{-0.2cm}{\hskip -0.0 true cm  $^3D_2$}}
\psfrag{xxx}{\raisebox{0.0cm}{\hskip -0.7 true cm  $E_{\rm lab}$ [GeV]}}
\psfrag{yyy}{\raisebox{0.0cm}{\hskip -0.3 true cm  $\delta$ [deg]}}
\parbox{7cm}{
\centerline{\hskip 0.3 true cm \psfig{file=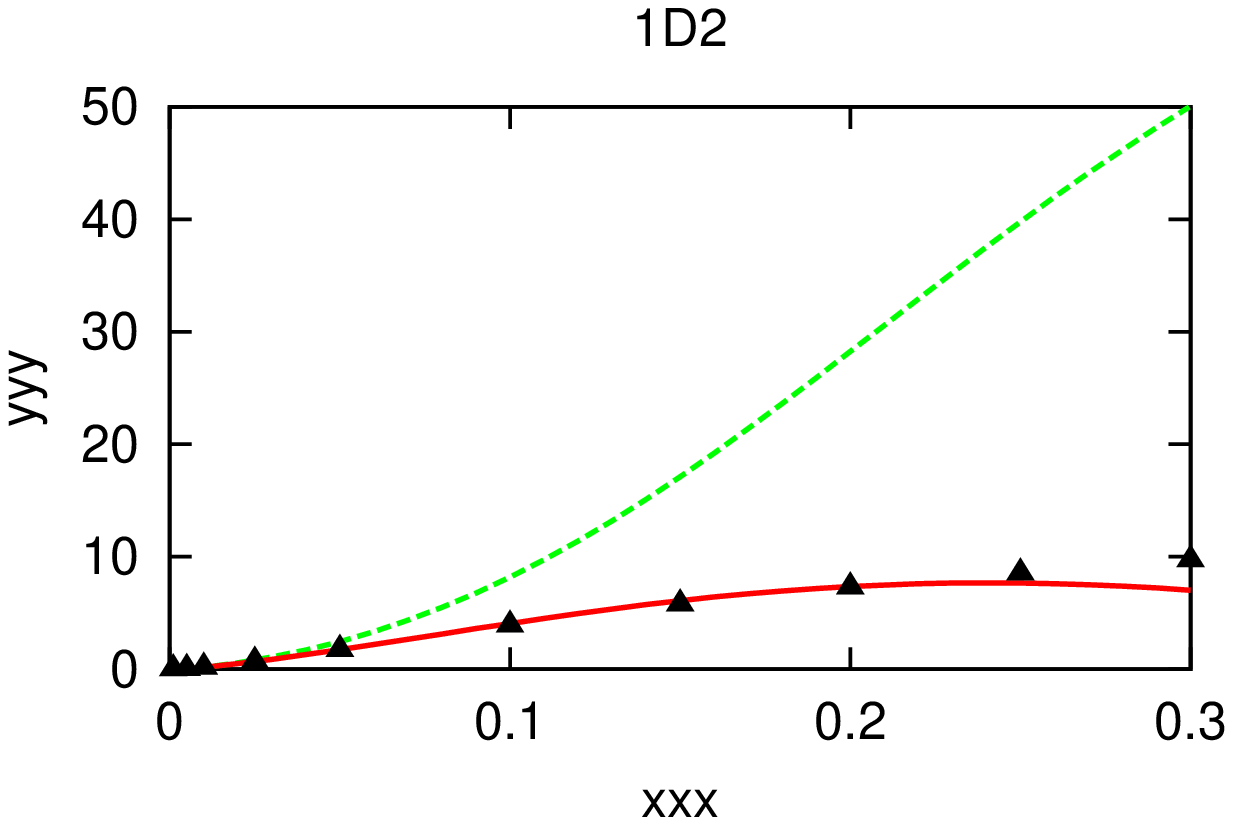,width=7.5cm,height=4.7cm}}}
\hskip 1 true cm
\parbox{7cm}{
\centerline{\hskip -0.0 true cm \psfig{file=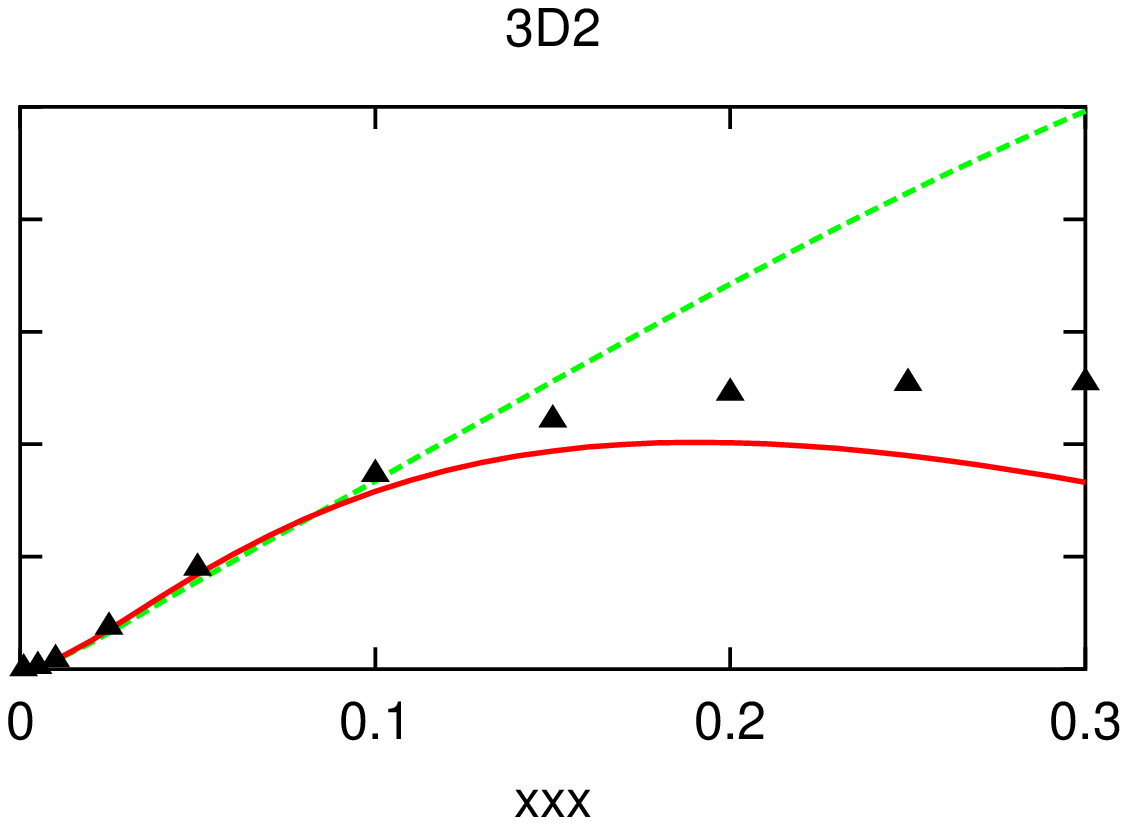,width=6.9cm,height=4.7cm}}}
\centerline{\parbox{14cm}{
\caption{\label{dwaves}$^1D_2$ and $^3D_2$ phase shifts at NNLO using 
the values of the $c_i$'s from ref.~\cite{Paul}.
The dashed lines show the Born approximation, whereas the solid lines correspond to the iterated solution 
with the exponential cut--off $\Lambda = 1000$ MeV.
The filled triangles are Nijmegen PSA results \cite{NPSA}}}}
\end{center}
\end{figure}
In fact, the NNLO results can serve as a counter example:
with the values of the $c_i$'s from eq.~(\ref{ci_central}), 
the Born approximation for the S--matrix, for instance, in the  $^1D_2$ partial wave deviates strongly
from the data already at $E_{\rm lab} \sim 100$ MeV, see fig.~\ref{dwaves}. Note that this result 
is parameter--free 
and cut--off independent.\footnote{Since we do not iterate the
potential,  we do not need to multiply it with the 
regulating function. Strictly speaking, of course, the EFT is only
defined with the cut-off procedure which would lead to the results
in Born approximation being multiplied with an overall factor. For
simplicity, we ignore this factor here. } 
Similar results have been published in ref.~\cite{KBW}.
On the other hand, as we showed in \cite{egm2},
taking the cut--off of the order of 1 GeV allows for a satisfactory 
description of all partial waves simultaneously. With such a large value of the cut--off, the central TPE potential  
becomes already so strongly attractive that unphysical deeply bound states appear in  the $D$--waves as well
as in the lower partial waves. 
Since the potential is very strong (and attractive) and 
there are no counter terms according to the power counting, changing the value of the cut--off clearly leads 
to strong variation of the $D$--wave phase shifts. This is illustrated and discussed 
in more detail in \cite{egm2}, \cite{EEdiss}.
\begin{figure}[hbt]
\begin{center}
\psfrag{1F3}{\raisebox{-0.2cm}{\hskip -0.0 true cm  $^1F_3$}}
\psfrag{3F3}{\raisebox{-0.2cm}{\hskip -0.0 true cm  $^3F_3$}}
\psfrag{xxx}{\raisebox{0.0cm}{\hskip -0.7 true cm  $E_{\rm lab}$ [GeV]}}
\psfrag{yyy}{\raisebox{0.0cm}{\hskip -0.3 true cm  $\delta$ [deg]}}
\parbox{7cm}{
\centerline{\hskip 0.3 true cm \psfig{file=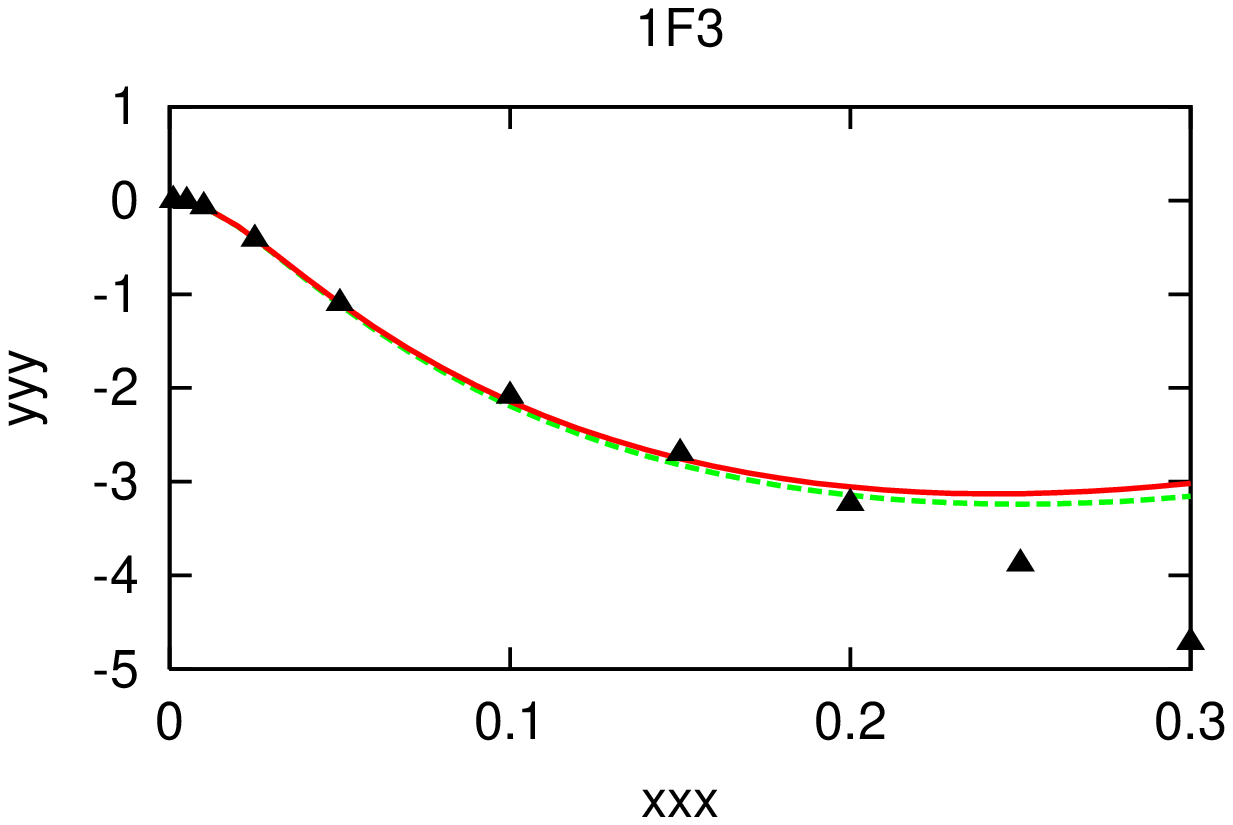,width=7.5cm,height=4.7cm}}}
\hskip 1 true cm
\parbox{7cm}{
\centerline{\hskip -0.0 true cm \psfig{file=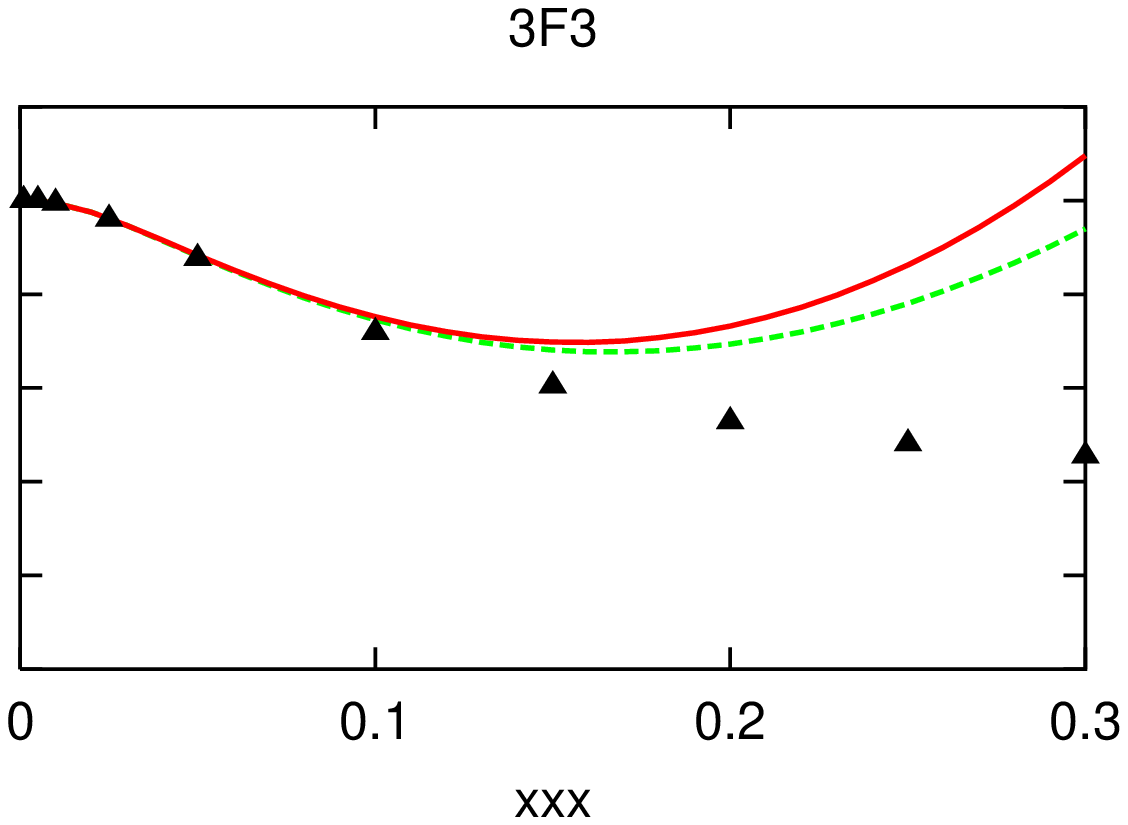,width=6.9cm,height=4.7cm}}}
\end{center}
\centerline{\parbox{14cm}{
\caption{\label{fwaves}$^1F_3$ and $^3F_3$ partial waves at NNLO
using the values of the $c_i$'s from ref.~\cite{Paul}.
For notations, see fig.~\ref{dwaves}.
The filled triangles are Nijmegen PSA results \cite{NPSA}}}}
\end{figure}
Note that this problem of the strong cut--off dependence does not show up in lower partial waves, where 
it is compensated by the cut--off dependence of the contact counter terms.
In the $F$--waves, where the potential
is already sufficiently weak (if the cut--off $\Lambda$ is chosen smaller or of the order of 1 GeV) 
and the Born approximation already does a good job, one has no problems with the cut--off dependence
as well.
This is shown in fig.~\ref{fwaves}. 
In spite of this fact one observes  
sizable deviations for most of the F--wave phase shifts from the Nijmegen PSA 
for energies larger than $E_{\rm lab} \sim 150$ MeV \cite{egm2}. Thus the only serious difficulty 
caused by the large values of the $c_i$'s in the NNLO analysis of the NN system is related to the cut--off 
dependence of the $D$--wave phase shifts.

The large numerical values of the $c_i$'s have also some consequences for three-- and more--nucleon systems, 
which will be discussed in detail in the next section. Here we only emphasize that effects 
from the inclusion of the 3N forces are expected to be much larger than in the standard 
scenario of nuclear physics. Note, however, that
the separate contributions of the 2N and 3N forces to 3N observables cannot be measured experimentally.

Let us now briefly summarize the consequences of the inclusion of the NNLO TPE with the large 
values of the $c_i$'s taken from the $Q^3$ analysis of the $\pi$N system \cite{Paul}:
\begin{itemize}
\item
First of all, including the subleading TPE allows for significant improvement in the description of
the low--energy observables in the NN system compared to NLO without introducing additional parameters, 
see ref.\cite{egm2} for more details. The phase shifts are mostly well reproduced.
\item
The central part of the potential shows a much stronger attraction
than  the one found in conventional models of the NN interaction \cite{KBW}. As a consequence, one 
has unphysical deeply bound states in the low NN partial waves.
\item
The predictions for D--waves depend on the cut--off. 
The optimal result is obtained for $\Lambda =1000$ MeV using the exponential regulator.
The potential projected onto the D--waves is strong and 
requires non-perturbative summation via the Lippmann--Schwinger equation.
The predictions for  F--waves deviate from the data at energies larger than $E_{\rm lab} \sim 150$ MeV.
In contrast, the peripheral waves are well described \cite{KBW}.
\item
One expects large effects from the 3NF.
\end{itemize}
Although the NNLO scenario dictated by the large values of the $c_i$'s differs strongly from our expectations
based on the experience with various 
phenomenological boson--exchange models, one cannot exclude this possibility a priori.
Indeed, the only serious problem with the large $c_i$'s is given by the 
strong cut--off dependence of the D--wave phase shifts.
However, this will probably not (or only weakly) affect chiral predictions
for experimentally measured quantities like the cross section, analyzing powers, etc. at low energies, where 
the contribution of the corresponding phases to physical observables is rather small. Further,
as already discussed in detail in ref.\cite{egm2}, at N$^3$LO it will be cured by dimension four
contact interactions. Furthermore the failure of the NNLO potential to describe various properties in 
the 3N and 4N systems does not yet indicate a problem, since we have not included the 3NF.
Because of the calculational difficulties in the treatment of the 3N and 4N systems in the 
presence of deeply bound states
it will take some time before all the implications of the chiral EFT at NNLO using
the large values of the $c_i$ will be explored in detail. These calculations need to be
done but will require a large amount of computing time. 

Having discussed consequences of the large values of the $c_i$'s for various properties of few--nucleon systems, 
we can ask ourselves, how confident we are, that the discussed scenario is indeed realized?
Several comments are in order:
\begin{itemize}
\item
First of all, we would like to stress the uncertainty in the determination of the $c_i$'s from 
$\pi$N scattering. 
The difference between the $c_i$'s from the  second and third order analyses of
$\pi$N scattering is considered to be an effect of third order, i.~e.~it should be suppressed by one power
of $Q$ compared to the second order values of the $c_i$'s. For that reason one can equally well take the $Q^2$--values
of the $c_i$'s in the NNLO analysis of the NN system, since the $c_i$'s enter only the NNLO and not the NLO
contribution to the effective potential. In principle, one can also take the values of the 
$\tilde{c}_i$ from the $Q^4$ analysis, which differ from the $c_i$'s by quark mass renormalizations of 
the order $M_\pi^2$. Taking different sets of the $c_i$'s from various analyses of the $\pi$N system, as 
described in the beginning of this section, might not cause significant variation in description 
of low--energy observables in the $\pi$N as well as NN systems, but lead to different scenarios.
\item
It is also possible that including higher order loop effects will reduce the strength of the central part of effective 
NN potential even if the $c_i$'s are numerically large.
\item
Finally, already at N$^3$LO one has to include new contact interactions with four derivatives, 
which also contribute in D--waves. These will not only reduce the cut--off dependence of the phase shifts, 
but may also provide additional repulsion and allow to avoid unphysical deeply bound states. 
The work by Entem and Machleidt \cite{entem}, who constructed a NN potential without deeply bound states by a
phenomenological extension of the NNLO chiral NN force,\footnote{To be precise, they   
included the N$^3$LO contact interactions 
and allowed for a partial wave dependent cut--off variation. Thus,
this extension is {\bf not} an EFT approach.} 
may serve as an indication of the importance of the higher order contact interactions.  
To ultimately clarify the situation one has to perform a complete analysis of the NN system at order N$^3$LO.  
\end{itemize}

It is interesting to understand the reason of (possibly) different scenarios in the EFT 
approach and in more phenomenological conventional boson-exchange (BE) models. 
It has been pointed out in ref.~\cite{BKM97} that the LECs $c_{3, 4}$ get
the dominant contributions from the intermediate $\Delta$ excitation. Also, the $\sigma$ and $\rho$ mesons
have been shown to play an important role in the saturation of the $c_i$'s. In particular, the constant
$c_1$ is completely saturated by the $\sigma$ \cite{BKM97}.
Let us now check whether these mechanisms of resonance saturation of the $c_i$'s are 
also realized in the OBE models of the NN interaction.
\begin{figure}[htb]
\vspace{1.2cm}
\centerline{
\psfig{file=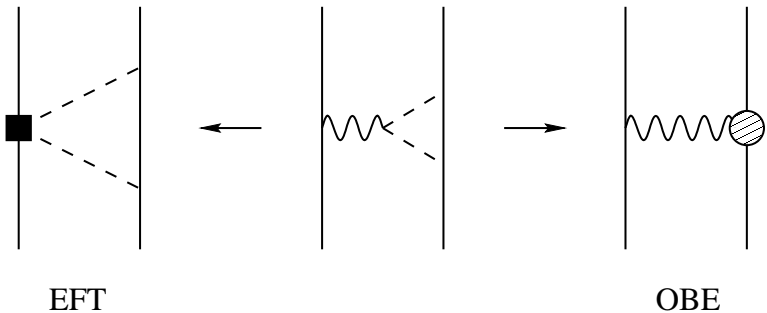,width=4in}}
\vspace{0.3cm}
\centerline{\parbox{14cm}{
\caption{\label{satur}  
Exchange of $\rho$--meson (wiggly line), which decays into two pions (dashed lines) 
and the corresponding diagrams in EFT (left--hand panel) and OBE models (right--hand panel).
The shaded blob represents the strong $\rho$N form factor in OBE models.
For remaining notations see figs.~\ref{fig1}, \ref{fig5}.
}}}
\vspace{0.7cm}
\end{figure}
While the resonance saturation of the $c_i$'s by heavy mesons can, in principle, be interpreted in terms
of OBE contributions as shown in fig.~\ref{satur}, where the pion loop in the graph 
in the middle of that figure 
contributes to the 
form factor of the corresponding heavy meson, the saturation by the $\Delta$ excitation cannot 
be represented in an appropriate way within the OBE models. Thus, a large portion of the subleading chiral TPE is 
absent in the conventional NN forces. 

A more detailed investigation of the two--pion exchange within the conventional 
many--boson exchange formalism gives rise to a better understanding of the reasons why the intermediate 
$\Delta$ plays only a modest role in the NN interaction. In the Bonn model of ref.\cite{Machl_rep}, 
which also allows for two--boson exchanges, one finds strongly attractive contributions from 
TPE. Note that this model also takes into account $\Delta$--excitations in the intermediate states.
The diagrams with intermediate $\Delta$--excitations 
have been shown to give the dominant contribution to the uncorrelated TPE.
While the TPE  model successfully describes high angular momentum partial waves, quantitative 
description of low partial waves appears to be impossible.  It is even stated 
in ref.~\cite{Machl_rep} that  ``the $2\pi$--contribution appears, in
general, too attractive and a consistent and quantitative description
of all phase shifts can never be reached''. It was shown that the strongly attractive contribution 
of the TPE in low partial waves is to a large extent canceled by the $\pi \rho$ diagrams. 
\begin{figure}[htb]
\vspace{1.2cm}
\centerline{
\psfig{file=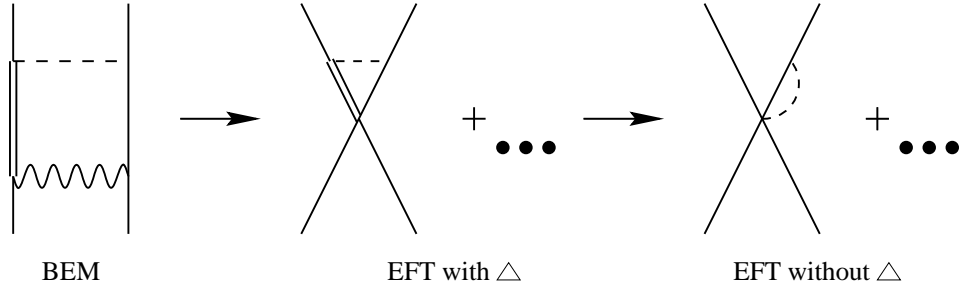,width=5.0in}}
\vspace{0.3cm}
\centerline{\parbox{14cm}{
\caption{\label{pirho}  
Representation of the $\pi \rho$ exchange diagram within EFT approaches with and without explicit 
$\Delta$. BEM stays for boson--exchange models. 
For notations see fig.~\ref{satur}.}}}
\vspace{0.7cm}
\end{figure}
The authors of ref.\cite{faassen} came to a similar conclusion.
The more detailed work on  correlated $\pi\rho$ exchange has 
been performed within the conventional formalism by Holinde and collaborators, see~\cite{Juelpirho}.
In fig.~\ref{pirho} we show one specific example of the $\pi \rho$ exchange with the corresponding
representation in the EFT approach. It is easy to see that the NLO\footnote{Note that if the $\Delta$--resonance 
is included explicitly via the ``small scale expansion'' \cite{Hemmert}, the strong attractive
central contribution to the TPE appears already at NLO and not at NNLO.}  
contribution to the effective 
potential from the diagram shown in fig.~\ref{pirho} only leads to renormalization of the corresponding LO contact 
interactions and thus will only influence the S--wave phase shifts. Thus one needs to go to higher 
orders beyond NNLO in the low--momentum expansion to see effects of the $\pi \rho$ exchange on the phase shifts in 
P-- and D--waves. The better way to observe the cancellation  between the $\pi \pi$ and $\pi \rho$ exchanges 
might be to include vector mesons as explicit degrees of freedom in the EFT.
That would however require a consistent power counting scheme, which has not yet
been constructed. 
\begin{figure}[htb]
\vskip 0.7 true cm
\centerline{
\epsfysize=7.1cm
\epsffile{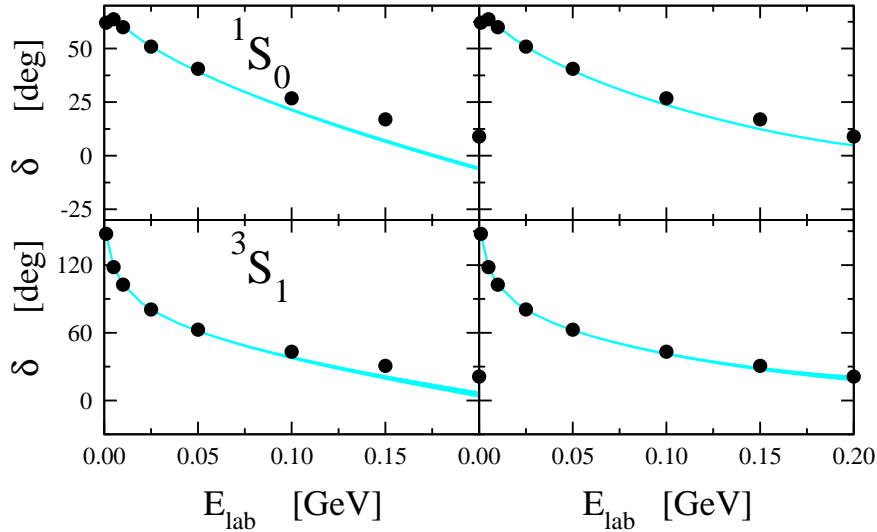}
}
\centerline{\parbox{14cm}{
\caption{\label{1s3s}Fits and predictions for the S-waves  for nucleon laboratory energies
$E_{\rm lab}$ below 200~MeV (0.2~GeV). Left/right panels: NLO/NNLO*
results. The cut-off is chosen between
500 and 600 MeV leading to the band. The
filled circles depict the Nijmegen PSA results \cite{NPSA}.}}}
\end{figure}

The study of the TPE within the Bonn model 
\cite{Hol78} also indicates 
a very important role of relativistic effects for diagrams with intermediate $\Delta$'s.
Incorporating relativistic corrections using IR regulated covariant baryon CHPT \cite{BeLe} within the 
EFT formalism has already been shown to reduce the strength of the subleading TPE by 
about 30\% \cite{Robil}.

\begin{figure}[htb]
\centerline{
\epsfysize=15.4cm
\epsffile{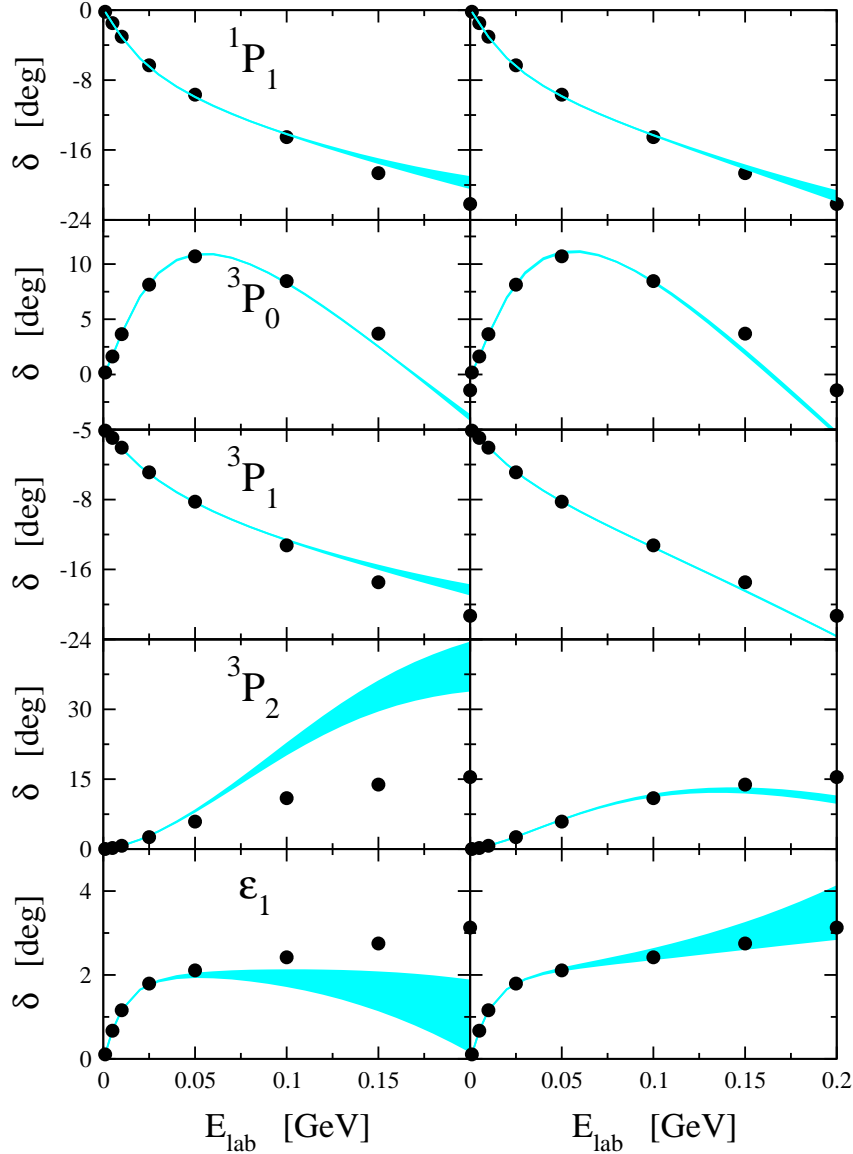}
}
\centerline{\parbox{14cm}{
\caption{\label{Pe}
Fits and predictions for the P-waves and the mixing parameter $\epsilon_1$  
for nucleon laboratory energies $E_{\rm lab}$ below 200~MeV (0.2~GeV). 
Left/right panels: NLO/NNLO* prediction. The cut-off is chosen between
500 and 600 MeV as shown by the band. The
filled circles depict the Nijmegen PSA results.}}}
\end{figure}

Although phenomenological boson--exchange models provide a plausible explanation of the 
fact that the $\Delta$--resonance does not play a significant role in NN scattering, 
additional model independent analysis is needed to improve on our understanding of the TPE.
In particular, more work on pion-nucleon scattering (dispersive versus chiral
representation), new dispersive analyses and more precise low-energy data
are needed to pin down these LECs to the precision required here.
\begin{figure}[htb]
\centerline{
\epsfysize=15.4cm
\epsffile{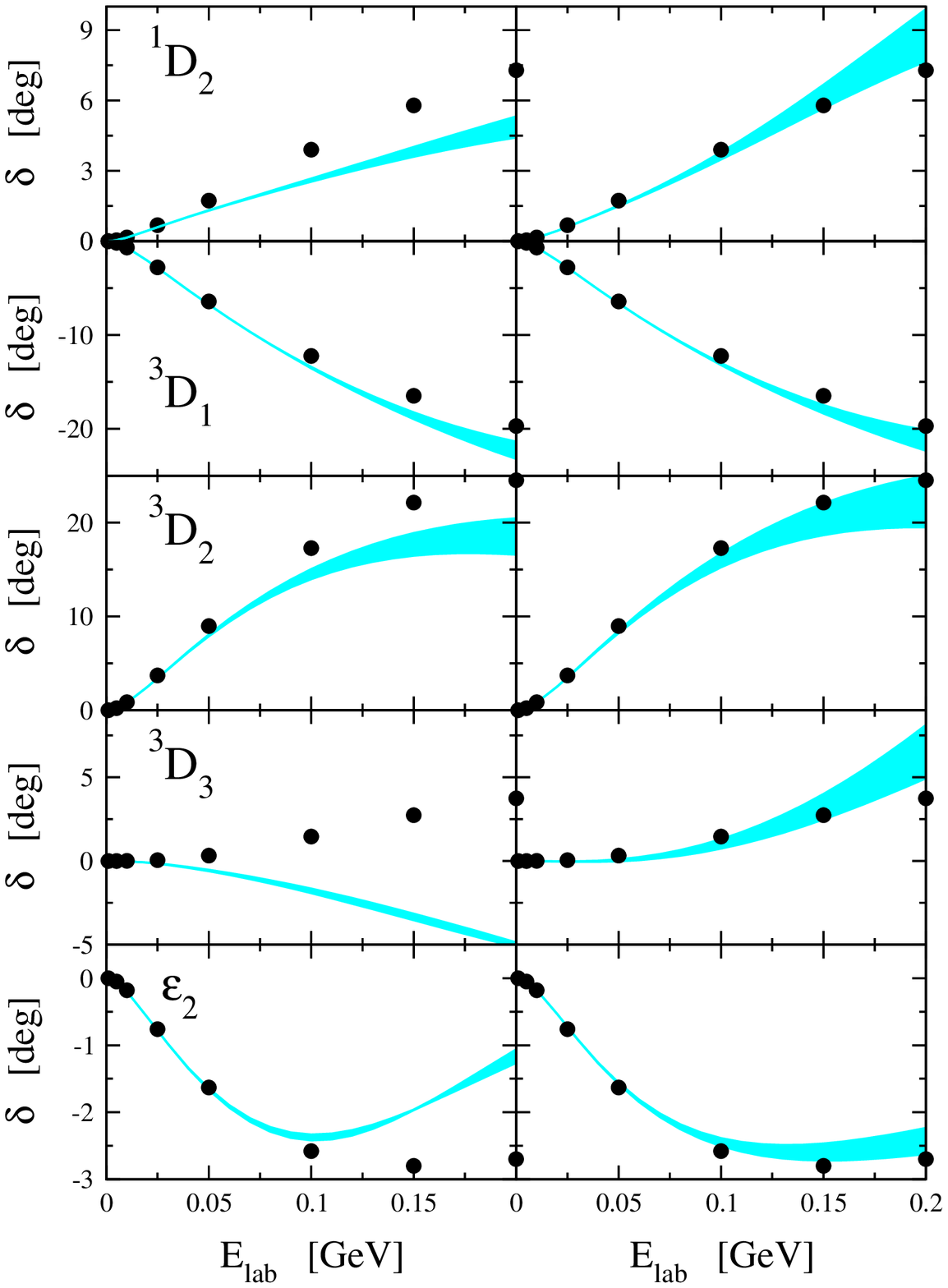}
}
\centerline{\parbox{14cm}{
\caption{\label{De}
Predictions for the D-waves and the mixing parameter $\epsilon_2$  
for nucleon laboratory energies $E_{\rm lab}$ below 200~MeV (0.2~GeV). 
Left/right panels: NLO/NNLO* prediction. The cut-off is chosen between
500 and 600 MeV as shown by the band. The 
filled circles depict the Nijmegen PSA results.}}}
\end{figure}

Motivated by the observed cancellation between the $\pi \pi$ and $\pi \rho$ exchanges and
by the fact that the $\Delta$ is not included as an explicit degree of freedom 
in existing OBE models and is supposed to play only a modest role for NN interactions at low energies,  
we constructed the NNLO* version of the effective potential \cite{Prag}, \cite{Resonance}, in which
we basically subtracted the $\Delta$--contributions from these LECs and allowed for some fine tuning.
This results in numerically reduced values of the $c_{3,4}$: 
\begin{equation}
\label{newci}
{c}_3 = -1.15 \,\mbox{GeV}^{-1}\;, \quad \quad c_4=1.20 \, \mbox{GeV}^{-1}\;.
\end{equation} 
As a consequence, the attraction of the central potential corresponding to chiral TPE
is reduced compared to the NNLO calculation of ref.~\cite{egm2}. Differently to the NNLO potential,
we also incorporated in the NNLO* version the leading isospin violating effect due to the pion mass 
differences in the OPE. 
\begin{figure}[htb]
\centerline{
\epsfysize=15.4cm
\epsffile{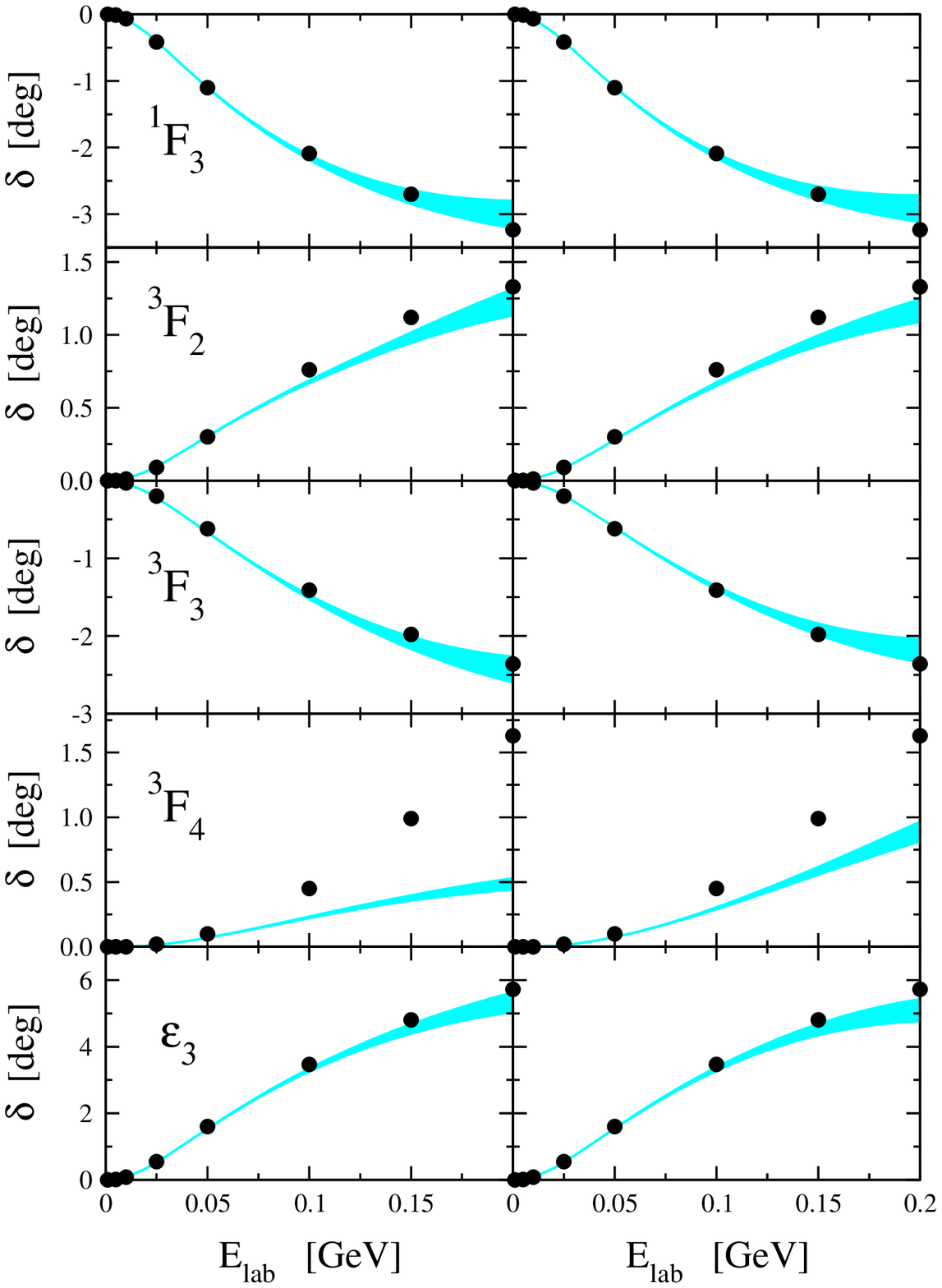}
}
\centerline{\parbox{14cm}{
\caption{\label{Fe}Predictions for the F-waves and the mixing parameter $\epsilon_3$  
for nucleon laboratory energies $E_{\rm lab}$ below 200~MeV (0.2~GeV). 
Left/right panels: NLO/NNLO* prediction. The cut-off is chosen between
500 and 600 MeV leading to the  band. The
filled circles depict the Nijmegen PSA results.}}}
\end{figure}

We are now in the position to discuss numerical results of the NNLO* potential.
First, we make some general remarks. For NLO (NNLO*), we fit to the Nijmegen S- and P-wave
phases and the $\epsilon_1$ mixing parameter up to $E_{\rm lab} = 50\,(100)\,$MeV.
These phase shifts at higher energies and for all higher partial waves are therefore
predictions. Throughout, we show the phase shifts
using the exponential regulator given in  eq.(\ref{reg1}). 
We are now able to use the same cut--off
range as we did at NLO.
Varying the cut-off $\Lambda$ between
500 and 600 MeV, we find a weakly changing $\chi^2/$ per degree of freedom. Also, for this range
of the cut-off we do not encounter any unphysical bound state in any partial wave,
which is in stark contrast to the NNLO results of \cite{egm2}. We note that one 
finds an increasing number of such deep bound states with increasing cut-off, eventually
leading to a limit cycle behavior (for details, see \cite{EMprep}). The theoretical
predictions at NLO and NNLO* for this cut-off range are indicated as bands in the
following figures. In most partial waves these bands get thinner when going
from NLO to NNLO* and are also visibly closer to the data (Nijmegen PSA). This is what one
expects from a converging EFT.

Let us now regard different partial waves. In fig.\ref{1s3s} we show 
the two S-waves. We find a
good description at NNLO* up to 200~MeV, which is comparable with (in case of the
$^1S_0$ partial wave slightly worse than) the NNLO results shown in ref.~\cite{egm2}.

\begin{figure}[htb]
\centerline{
\epsfysize=15.4cm
\epsffile{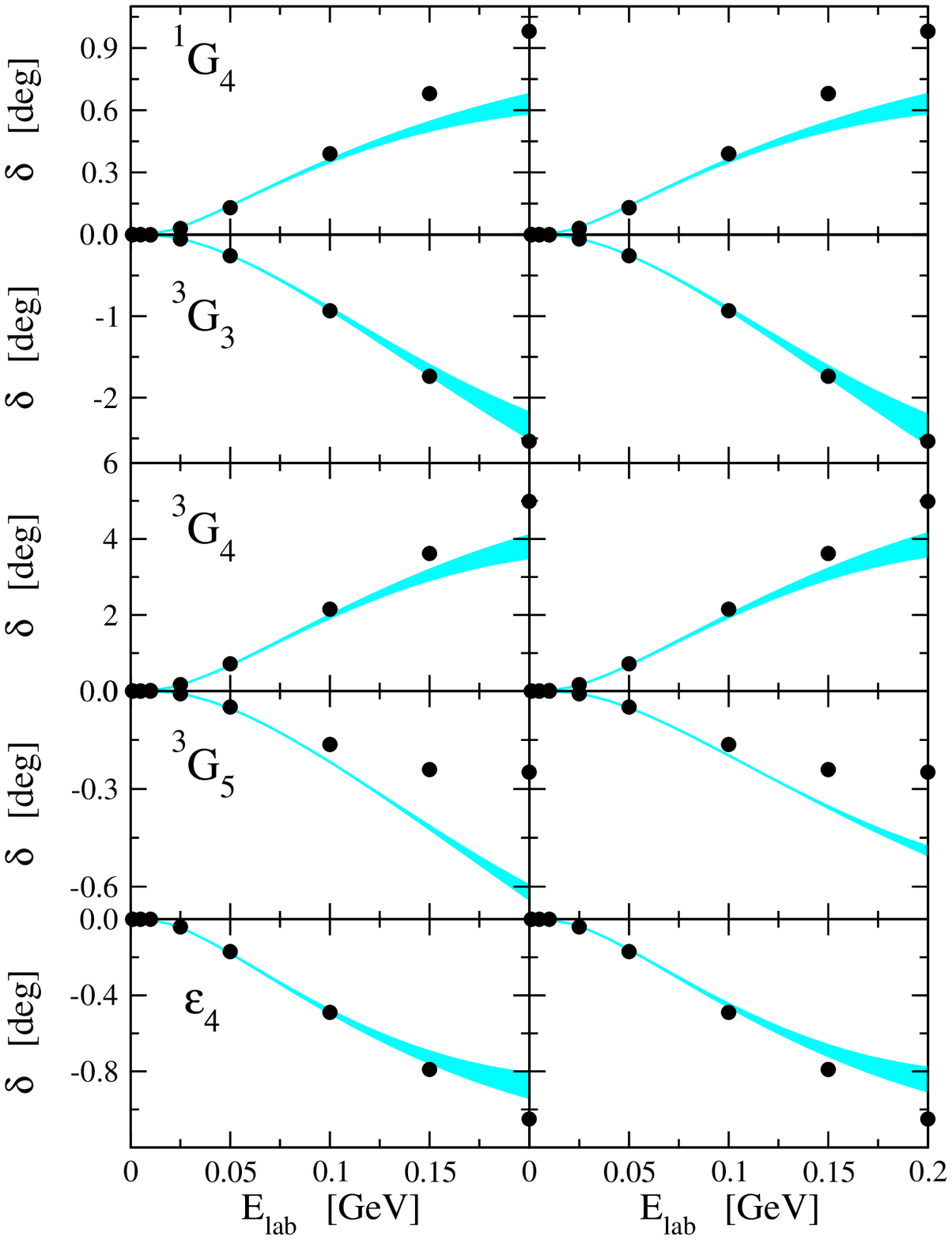}
}
\centerline{\parbox{14cm}{
\caption{\label{Gs}Predictions for the G-waves and the mixing parameter $\epsilon_4$  
for nucleon laboratory energies $E_{\rm lab}$ below 200~MeV (0.2~GeV). 
Left/right panels: NLO/NNLO* prediction. The cut-off is chosen between
500 and 600 MeV leading to the  band. The 
filled circles depict the Nijmegen PSA results.}}}
\end{figure}

Consider next the P--waves and the mixing angle $\epsilon_1$ shown in fig.\ref{Pe}.
The most visible improvement from NLO to NNLO* is observed for $^3P_2$ and
$\epsilon_1$. We also note that the description of $^3P_2$ is better than  
in the NNLO case shown in ref.\cite{egm2}.
While the NNLO corrections to the NLO results for the $^1P_1$, $^3P_1$ and $^3P_2$ 
partial waves (see fig. 5 in ref.~\cite{egm2}) go in the right directions, 
the observed effects turn out to be too large and lead to significant deviations
from the data. This is cured in the NNLO* version, as can be seen from  fig.\ref{Pe}.
The NNLO* and NNLO results for the  $^3P_0$ partial wave are very similar 
to each other and to the NLO calculation. 

Let us now discuss the D--waves and the mixing angle $\epsilon_2$.
These are of particular interest since at NNLO* no parameters
enter and we already discussed the strong cut-off sensitivity found at
NNLO. As shown in fig.~\ref{De}, this cut--off sensitivity is sizeably
reduced at NNLO* (in comparison to NNLO) 
and one obtains an overall good description of all
D--waves up to laboratory energies of about 200~MeV. We remark that the
important $\pi\pi$ correlations which are at the heart of the dramatic
improvement in $^3D_3$ from NLO to NNLO* are still present (as in
NNLO) since they are driven by the physics behind the LEC $c_1$. 
Note also the significant improvement for the $\epsilon_2$.

The NNLO* corrections get weaker for F-- and higher
partial waves. In contrast to the strong NNLO effects in the 
F--waves, which cause significant deviations of the phase shifts
for the results of the Nijmegen PSA, 
the NNLO* results can be viewed as small corrections to the NLO
calculations, see fig.~\ref{Fe}. Indeed, in most cases the difference 
between the NLO and NNLO* predictions is very small. The only exception
is observed for the $^3F_4$ partial wave. Here the NNLO* corrections go in the 
right direction but are still not sufficient to reproduce the 
phase shift appropriately at energies larger than $50-100$ MeV.

The peripheral partial waves (G,H,I, $\ldots$) are mostly well
described. Most of these are dominated by OPE.
However, in very few cases the large values
of the $c_i$'s  
were needed to bring the prediction in agreement
with the data, see refs.\cite{KBW,egm2}. In the NNLO* potential,
the weakened TPE does not provide enough strength as e.g. seen
in $^3G_5$, cf. fig.\ref{Gs}. Similar remarks hold for the $H$ and
$I$ phase shifts; we refrain from showing these here.

We now turn to the bound state (deuteron) properties. We have not
fine-tuned the parameters to exactly reproduce the binding energy.
It is already described within 2\% for the range of cut-offs
considered here. In table~\ref{tab:D} we collect the deuteron
properties at NLO and NNLO* (for $\Lambda = 500$ and $600$~MeV) in
comparison to the NNLO results (obtained with an exponential regulator
with $\Lambda = 1.05$~GeV) and the CD-Bonn potential (as one generic
high-precision potential). Most deuteron properties are well reproduced
and improve when going from NLO to NNLO*. 
We also note that all NNLO* predictions (except the one for the 
quadrupole moment) are between the 
NLO and NNLO results.\footnote{One should keep in mind, that while the 
NLO and NNLO* results are given within the theoretical uncertainty, which corresponds
to a cut--off variation, the results at NNLO are only shown for the optimal choice   
of the cut--off $\Lambda= 1050$ MeV.} The quadrupole moment is only slightly improved at NNLO*, 
while the NNLO correction for this quantity goes in the wrong direction. One, however, 
still observes a discrepancy of about 7\% to the experimentally observed value 
(see, however, the recent discussion by Phillips \cite{PhilPrag} why this 
failure is not unexpected). It has also been noted in \cite{walzl} that fine-tuning
the binding energy can slightly improve the prediction for $Q_d$.
The NNLO* and NNLO corrections go in the wrong (right) direction for the asymptotic
$D / S$ ratio $\eta$ (the asymptotic S--wave normalization $A_S$). The improvement 
for $A_S$ at NNLO* is significant compared to NLO but still leaves 
space for N$^3$LO corrections. 
The same holds true for the root--mean--square matter radius $r_d$. 
We note that the (unobservable) D-state probability
is reduced as compared to the NNLO result and agrees more with the one found
using CD-Bonn potential.

\renewcommand{\arraystretch}{1.2}    

\begin{table}[htb] 
\begin{center}

\vskip 0.6 true cm 
\begin{tabular}{||l||c|c||c|c||c||c||c||}
    \hline
    & \multicolumn{2}{|c||}{NLO}  &  \multicolumn{2}{|c||}{NNLO*}  &  &   &     \\ 
    \cline{2-5}   
    & 500 MeV  & 600 MeV & 500 MeV & 600 MeV & \rb{NNLO} & \rb{CD-Bonn} & \rb{Exp.} \\
\hline  \hline
$E_d$ [MeV] & $-$2.152     & $-$2.165 & $-$2.182 & $-$2.189  & $-$2.224  & $-$2.225 & $-$2.225 \\
    \hline
$Q_d$ [fm$^2$] & 0.265      & 0.266     & 0.265    & 0.268     & 0.262     & 0.270 & 0.286 \\
    \hline
$\eta$ & 0.0248             & 0.0248    & 0.0247   & 0.0247    & 0.0245    & 0.0255 & 0.0256\\
    \hline
$r_d$ [fm] & 1.975          & 1.975     & 1.970    & 1.969     & 1.967     & 1.966 & 1.967 \\
    \hline
$A_S$ [fm$^{-1/2}$] & 0.862 & 0.866     & 0.871    & 0.874     & 0.884     & 0.885 & 0.885\\
    \hline
$P_D [\%]$ &   3.17         & 3.62      & 3.65     & 4.52      & 6.11      & 4.83 &  -- \\
    \hline
  \end{tabular}
\caption{Deuteron properties derived from our chiral potential
    at NLO and NNLO* (for the cut-off range considered throughout)
    compared to the NNLO results of \protect\cite{egm2},
    one ``realistic'' potential and the data. Here, $E_d$ is the
    binding energy, $Q_d$ the quadrupole moment, $\eta$ the asymptotic
    $D/S$ ratio, $r_d$  the root--mean--square matter radius, $A_S$ the 
    strength of the asymptotic S--wave normalization and $P_D$ the D-state 
    probability.
\label{tab:D}}
\end{center}
\end{table}

\begin{figure}[htb]
\centerline{
\epsfysize=10cm
\epsffile{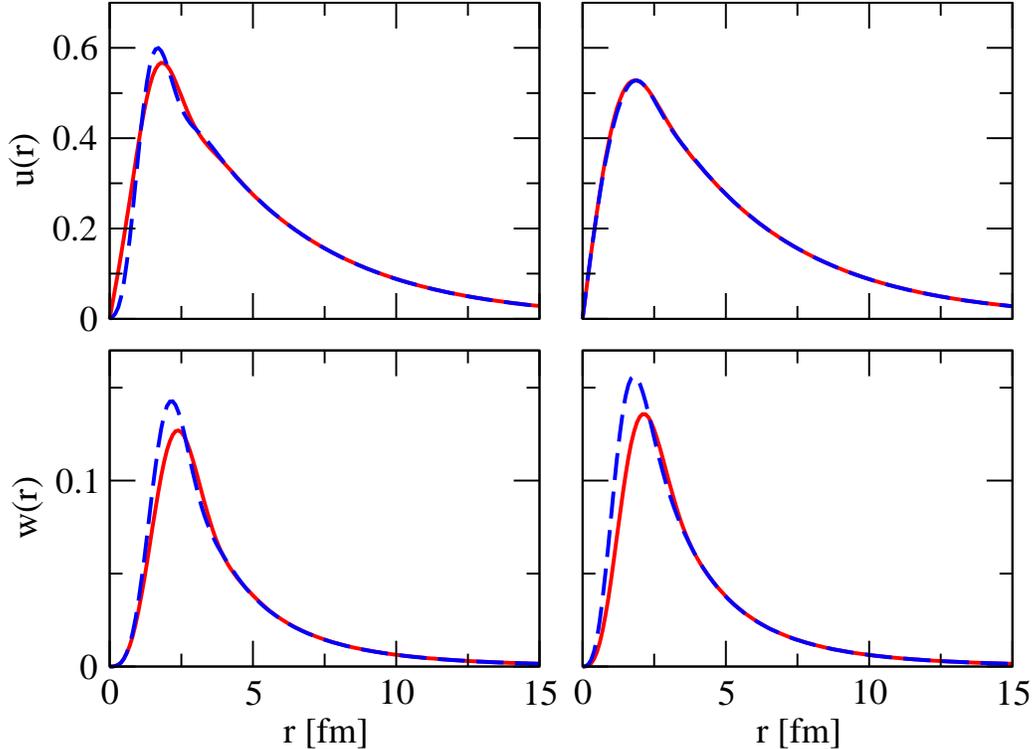}
}
\centerline{\parbox{14cm}{
\caption{\label{deut}Deuteron wave functions at NLO (left panels) and 
NNLO* (right panels) in coordinate space. The solid (dashed) lines
correspond to the cut--off $\Lambda=500$ MeV ($\Lambda=600$ MeV).}}}
\end{figure}

The NNLO* deuteron coordinate space S- and D-wave functions $u(r)$ and $w(r)$,
respectively, are shown in fig.\ref{deut}. By construction, they have no nodes
and agree quite well with e.g. the CD-Bonn wave functions. This lets one expect 
that the NNLO* potential when applied to the 3N and 4N systems gives results closer
to calculations based on conventional potentials as does NNLO. 
We will discuss this issue in the following two sections.

Let us now summarize the presented numerical findings for the 2N low--energy observables.
Altogether it can be seen that the NNLO* potential leads to results, which are 
significantly improved compared to the NLO ones and allows for a quantitatively 
rather good description of the $np$ phase shifts up to $E_{\rm lab} \sim 200$ MeV. 
While the results for observables at NNLO* and NNLO seem to be of comparable quality and
in many cases do not significantly differ from each other, these two versions of the chiral potential 
suggest quite different scenarios, as discussed above. 
It is difficult to give  preference  
to the NNLO* or the NNLO version of the chiral potential.
In principle, the LECs $c_i$ should be taken from the analysis of $\pi$N 
scattering and no readjustment should occur, if sufficiently many
terms of the chiral expansion of the $\pi$N scattering amplitude
appear in the TPE potential and the $\pi$N parameters are
precisely known. However, with the presently available best
determinations of these LECs at $Q^2$ and $Q^3$, one gets 
a very strong attractive central part of the TPE and, as a consequence,
encounters unphysical deeply
bound virtual states. Further, they lead to an unconventional balance between
two-- and many--nucleon forces in systems with three (or more)
nucleons. On the other hand, boson--exchange phenomenology clearly
indicates the suppression of contributions with delta intermediate
states based on cancellations with e.g. $\pi\rho$ exchanges. Such a
scenario is realized in the NNLO* potential, which does not lead
to unphysical bound states in the NN system for reasonable choices of
the cut-off. Progress can come from different directions: Further
investigations  of the $\pi$N system
at higher orders in chiral expansion as well as new data
(eventually combined in dispersion relations) may allow for more precise determination
of the $c_i$'s, so that one would be able to discriminate the physically relevant scenarios 
of the NN interaction. On the other hand, the final word on the choice
of regulator is not yet spoken - one may still contemplate the
construction of a coordinate-space regulator that modifies the TPE at
short distances such that no unphysical bound states appear. At present,
this is only a speculation (we refer to \cite{bbsk} for some related work).
Clearly, more work in this direction is mandatory. 
For the time being we consider it legitimate to use
the NNLO* potential in applications to the 3N and 4N systems. For the sake of completeness,
we will, however, briefly discuss in the next section NNLO predictions for the 3N system, 
before we switch to the central issue of this paper and present the NNLO* results 
for 3N and 4N systems.

\section{NNLO Predictions for the 3N System}
\setcounter{equation}{0}

As has been shown in \cite{egm2} the NNLO NN forces describe the Nijmegen NN
phase shift values significantly better than the NLO ones. 
We would like to remind the reader
that there occur spurious bound states in $S$--, $P$-- 
and $D$--waves, as already mentioned in the preceeding section. As
a consequence of these deeply bound states, the deuteron
wave function at NNLO  has nodes below about 1~fm, which are not present at NLO 
(and NNLO*) or
using conventional NN forces. In agreement with the correct description of
the low--energy $^3S_1 - ^3D_1$ phase shift parameters those nodes also do not
influence the low--energy deuteron properties: its binding energy, the asymptotic D/S
ratio, the root--mean--square matter radius, the asymptotic S--wave
normalization constant and the quadrupole moment, which are in good to fair
agreement with the experimental values.
\begin{figure}[htb]
\vspace{1.2cm}
\centerline{
\psfig{file=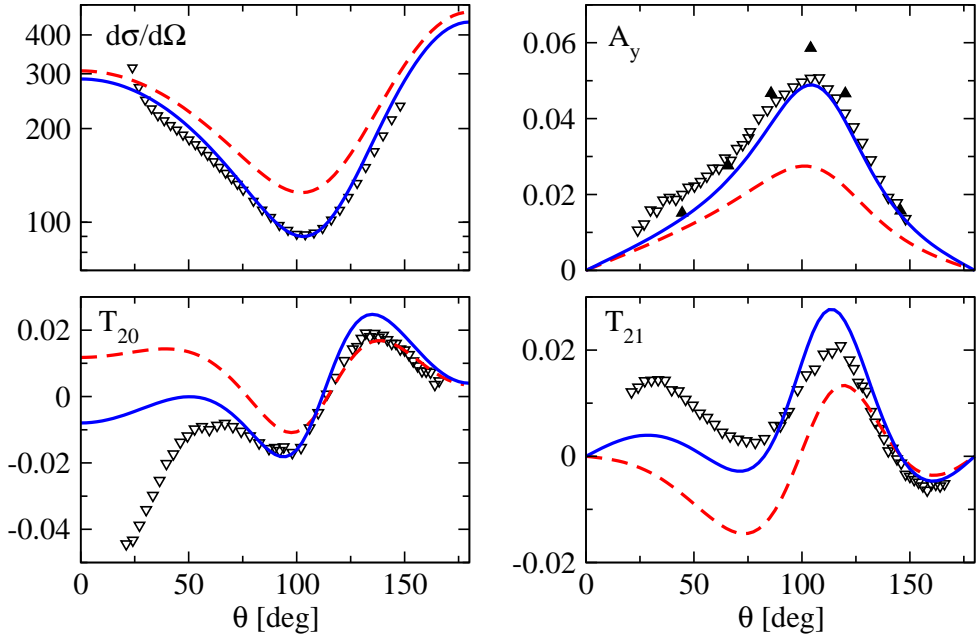,width=5.3in}}
\vspace{0.3cm}
\centerline{\parbox{14cm}{
\caption{\label{fignnlo}  
Differential cross section (in [mb/sr]) and analyzing powers $A_y$, $T_{20}$ and $T_{21}$ for 
elastic {\it nd} scattering at $E_{\rm lab}=$3 MeV. The solid and dashed lines correspond to 
predictions based on the CD-Bonn and NNLO potentials. The open (filled) triangles are 
{\it pd} \cite{Sag94}, \cite{Shim95} ({\it nd} \cite{McA85}) data.}}}
\vspace{0.7cm}
\end{figure}

In turning to  the  3N system one encounters in the Faddeev formulation NN
t--matrices which are taken off the energy--shell. The energy argument is
\begin{equation}
E_2 = E- \frac{3}{4m} q^2~,
\end{equation}
where $E$ is the fixed 3N energy and $(3/4m) q^2$ the
varying kinetic energy of the third particle in relation to the pair of
nucleons interacting via the NN t--matrix. Since $q$ varies between 0 and infinity one
necessarily hits the spurious bound state energies, which occur as poles
of the NN t--operator. Physically spoken this has the consequence that
the normal 3N bound state is not stable but decays into two fragments,
a deeply bound spurious NN bound state and a nucleon. In practice this decay is
rather weak, however, and can be neglected since the physical 3N
bound state has little overlap with the short ranged spurious NN bound state.
In addition, one
has to expect that there will be spurious 3N bound states at extremely
large negative energies in the GeV region.
Calculating the 3N observables in the presence of deeply bound spurious states in
the NN system requires some precautions. Of course, the ultimate way to calculate 
3N observables in the presence of deeply bound NN states would be 
to treat the poles of the NN t--operator explicitly in the corresponding integral 
equation. The much easier approximate way is to restrict
the virtual $q$--values such that  $E_2$ does not reach the energies of the
spurious bound states, which are in the GeV region. 
Alternatively,  one can transform
the two--body Hamiltonian  in such a manner  that the NN phases do not change but the
spurious bound state energies are moved towards high positive energies,
where they cause no technical obstacles. This can be achieved for
instance by the following simple change of the two--nucleon force 
(starting from the Hamiltonian $H=H_0+V$)
\begin{equation}
\tilde{V} = V + \sum_i |\Psi_i\rangle \alpha_i \langle \Psi_i | 
\end{equation}
leading to the modified Hamiltonian $\tilde{H} = H_0 + \tilde{V}$ with
shifted eigenvalues corresponding to spurious eigenstates,
\begin{equation}
\tilde{H}  |\Psi_i\rangle  = (E_i+\alpha_i)  |\Psi_i\rangle~,
\end{equation}
where the $\alpha_i$  are sufficiently large positive energies, $|\Psi\rangle_i$ the 
spurious bound states and $E_i$ the binding energy of the spurious state  $|\Psi_i\rangle$. 
Note that such a projection does not influence 
the 2N phase shifts. Also the deuteron wave
function remains unchanged.

In this work we do not aim to apply the NNLO potential to the 3N and 4N systems and 
only want to demonstrate that one needs strong 3NFs to describe the data. The approximate 
methods described above are therefore sufficient for our present purpose.
We solved the 3N Faddeev equation for the triton using $\tilde{V}$ instead of $V$
and this for NNLO. With the cut--off $\Lambda=1000$ MeV in the NN system 
we found for triton binding energy $E=-3.8$ MeV. This number turned out to be
nearly independent of the actual values of $\alpha_i$ (which are of the order of a  
few GeV). 
A very close value arises if on sticks to the original NN force $V$ at NNLO and restricts the
range of $q$ values as mentioned above. One has to conclude that this form
of the NN force requires strong 3N forces to account for the missing
binding energy. Notice, however, that these required 3N forces may still be 
much weaker than the corresponding 2N ones. Indeed, we found an expectation 
value for the potential energy in the triton at NNLO of about $-$172 MeV, 
which is much larger than the one observed for various 
high--precision potentials of the order $-$40 to $-$50 MeV.\footnote{Thus, 
the missing binding energy of about 4 MeV  
for the triton to be provided by 3N forces 
is still much smaller compared to the strength of the 
2N interaction.}
The corresponding large value for the kinetic 
energy has, in principle, to be expected due to the additional nodes in the 
deuteron and triton wave functions in the short distance range, which are caused by
the deeply bound states.
 
In view of the results for the triton binding energy
one also has to expect that theoretical
3N scattering observables based only on the NNLO NN force (i.e. neglecting the 
3NFs) will be in
conflict with the data. This is indeed the case as shown in fig.\ref{fignnlo} for a
few examples.

Thus, we conclude that taking into account only the 2N interaction at NNLO and
neglecting the corresponding 3NFs does not allow for a correct description 
of the 3N observables. This presumably will be corrected by the inclusion of the 
3N  forces, which because of consistency in the power counting has to be taken into
account at NNLO. It will be interesting in the future to check this statement explicitely.

\section{3N and 4N Predictions with the NNLO* NN Potential}

We use the Faddeev-Yakubovsky scheme to solve for the 3N and 4N bound
states and the 3N scattering observables as described in \cite{KaG92},\cite{GloeckleREP}. The
calculations are fully converged with respect to the number of partial
wave states and standard numerical discretizations. Table~\ref{tab:bind} shows the
results for the 3N and 4N binding energies using the  NLO and the NNLO* NN
potentials. Note that for the NLO version the numbers slightly different from the ones
published in \cite{ourPRL} appear since we have now taken into account the leading 
isospin violating effect due to the charged to neutral pion mass difference 
in the OPE.
We see a clear reduction of the cut-off dependence in going from
NLO to NNLO*, as it is expected from a converging EFT. For reasons of comparison, we also
display the kinetic energy and the probabilities of the various
ground state components ($S,P,D$) in $^4$He.
\begin{table}[htb]
\vskip 0.6 true cm 
\begin{center}
\begin{tabular}[t]{|l||r||r|r|r|r|r|}
\hline
Potential & $E(^3{\rm H})$  &  $E(^4\rm{He})$ &   $T$   
          & $S$ [\%] & $P$ [\%] & $D$ [\%] \\
\hline
\hline
NLO, 500     &  $-$8.544 & $-$29.57   & 61.4   & 94.71   & 0.07  &  5.22   \\
NLO, 600     &  $-$7.530 & $-$23.87   & 77.6   & 92.60   & 0.11  &  7.29   \\
\hline
NNLO*, 500   &  $-$8.590 & $-$29.96   & 62.2   & 93.65   & 0.10  &  6.25   \\
NNLO*, 600   &  $-$8.245 & $-$27.87   & 64.9   & 90.61   & 0.17  &  9.22   \\
\hline\hline
AV-18        &  $-$7.628 & $-$24.99   & 97.8   & 85.89   & 0.35  &  13.76  \\ 
CD-Bonn      &  $-$8.013 & $-$27.05   & 77.2   & 89.06   & 0.22  &  10.72  \\ 
\hline\hline                      
exp          &  $-$8.48  & $-$29.00   & ---    &  ---    & ---   &  ---    \\
\hline
\end{tabular}
\vskip 0.3 true cm
\caption{\label{tab:bind} Theoretical $^3$H and $^4$He binding energies for
different cut-offs $\Lambda$ at NLO and NNLO*  
compared to the AV-18 and CD-Bonn predictions (point Coulomb interaction 
perturbatively removed), the experimental 
$^3$H binding energy and 
the Coulomb corrected $^4$He binding energy in MeV.  
The kinetic energies $T$ (in MeV) and $S$, $P$ and $D$ 
state probabilities for $^4$He are also shown.}     
\end{center}
\end{table}
The resulting binding energies for NNLO* are near the
experimental data and larger than the values typically achieved with
conventional potentials. The results for two representatives, AV18 and
CD-Bonn, are also displayed in Table~\ref{tab:bind}. 
Note that the NNLO* results encompass the experimental values, quite in contrast to
the realistic potentials.
We remark, however, that the
chiral NN forces employed up to now are for the $np$ system and therefore do
not yet take all relevant isospin violating effects 
into account \footnote{Such effects can be dealt with
in nuclear EFT as discussed e.g. in \cite{WME}.}. Experience tells us that this leads to an
unphysical increase in the binding energy of about 200~keV (1~MeV) in $^3$H 
($^4$He). Nevertheless in relation to conventional forces one ends up
close to the experimental data for $^3$H and $^4$He using  the NNLO* NN potential
and consequently will need  smaller contributions of 3N forces 
than using conventional NN forces. 
\begin{figure}[htb]
\centerline{
\epsfysize=12.5cm
\epsffile{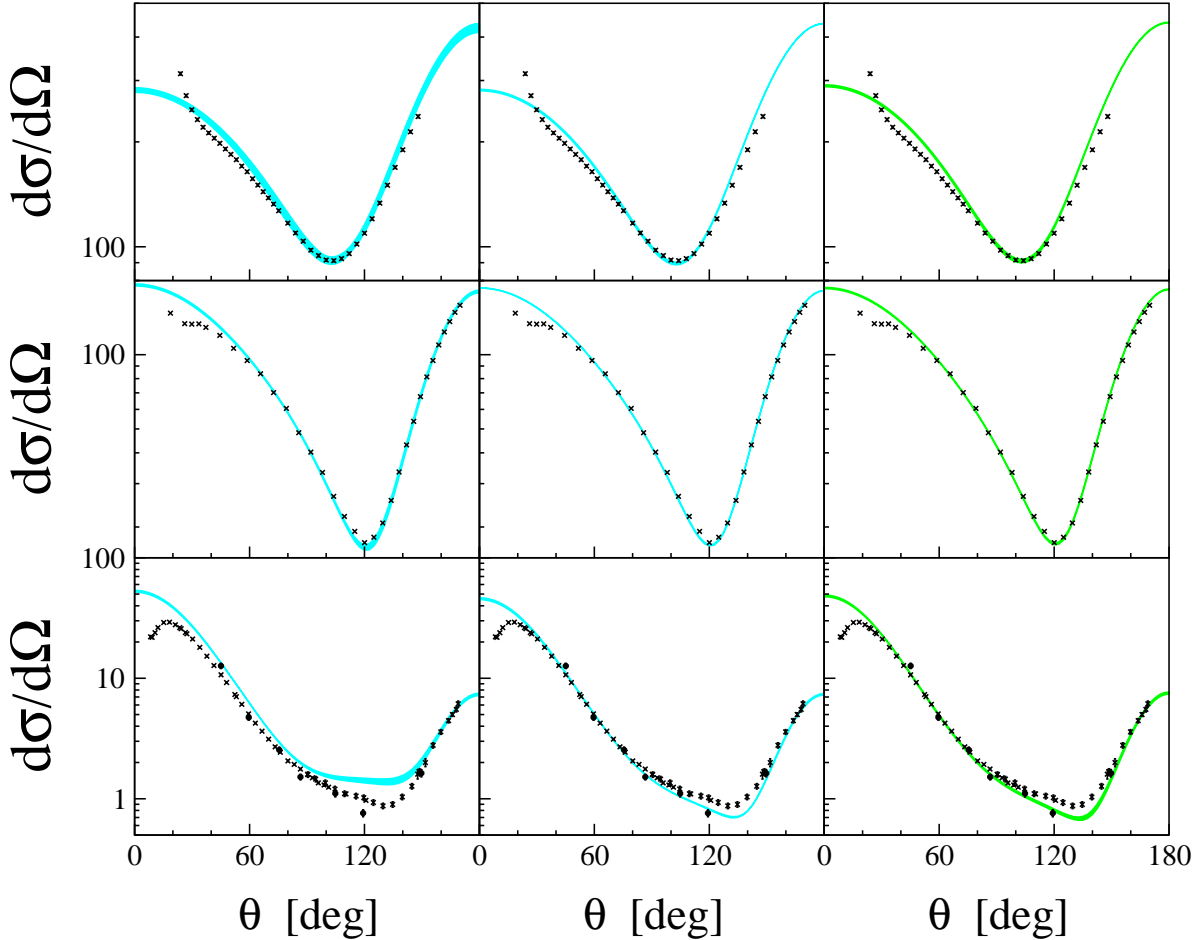}
}
\centerline{\parbox{14cm}{
\caption{\label{fig:sig}Differential cross section for elastic $nd$
scattering in [mb/sr] for $E_{\rm lab} = 3,10,65\,$MeV (top to bottom). 
Results at NLO (left panel) and NNLO* (middle
panel). The bands correspond to the range $\Lambda = 500$ to $600$ MeV.
Results based on the high-precision potentials (CD-Bonn, AV-18, 
Nijm-93, Nijm-I,II) are shown in the right panel.
Here the bands refer to the spread in predictions using the various potentials.
The crosses are $pd$ data: at 3 MeV from \cite{Sag94}, at 10 MeV from \cite{Sper84},
and at 65 MeV from \cite{Shim82}. The circles at 65 MeV are $nd$ data from \cite{Rue91}.
}}}
\end{figure}

\begin{figure}[htb]
\centerline{
\epsfysize=12.5cm
\epsffile{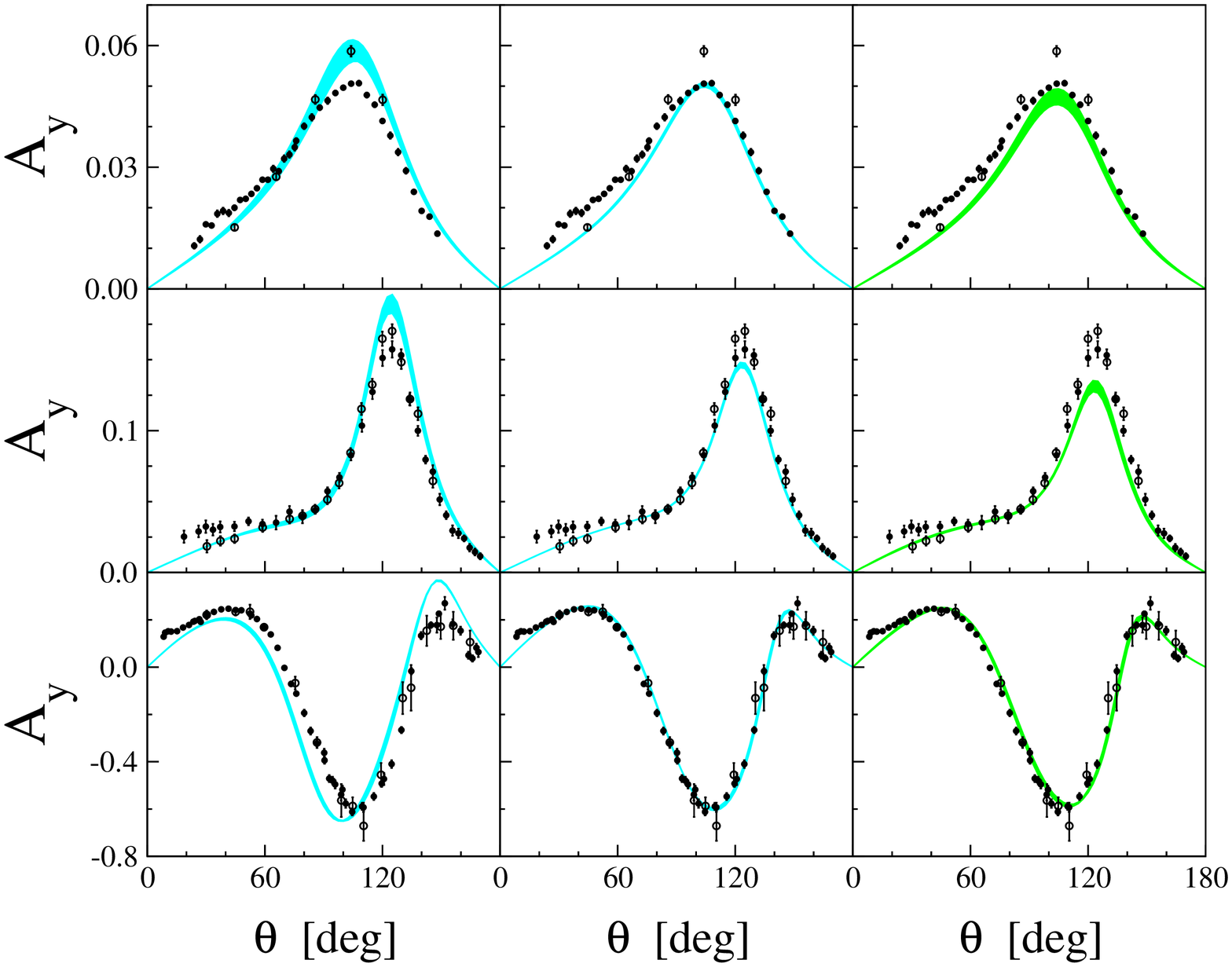}
}
\centerline{\parbox{14cm}{
\caption{\label{fig:ay}Analyzing power $A_y$ for elastic $nd$
scattering, for $E_{\rm lab} = 3,10,65\,$MeV (top to bottom).
Data at 3 MeV are from \cite{McA85} ($nd$, open circles) and 
\cite{Shim95} ($pd$, filled circles),
at 10 MeV
from \cite{Torn82} ($nd$, open circles) and \cite{Sper84} ($pd$, filled circles), 
and at 65 MeV from \cite{Rue91}
($nd$, open circles) and \cite{Shim82} ($pd$, filled circles).
For further notations, see fig.\ref{fig:sig}.
}}}
\end{figure}

\begin{figure}[htb]
\centerline{
\epsfysize=12.5cm
\epsffile{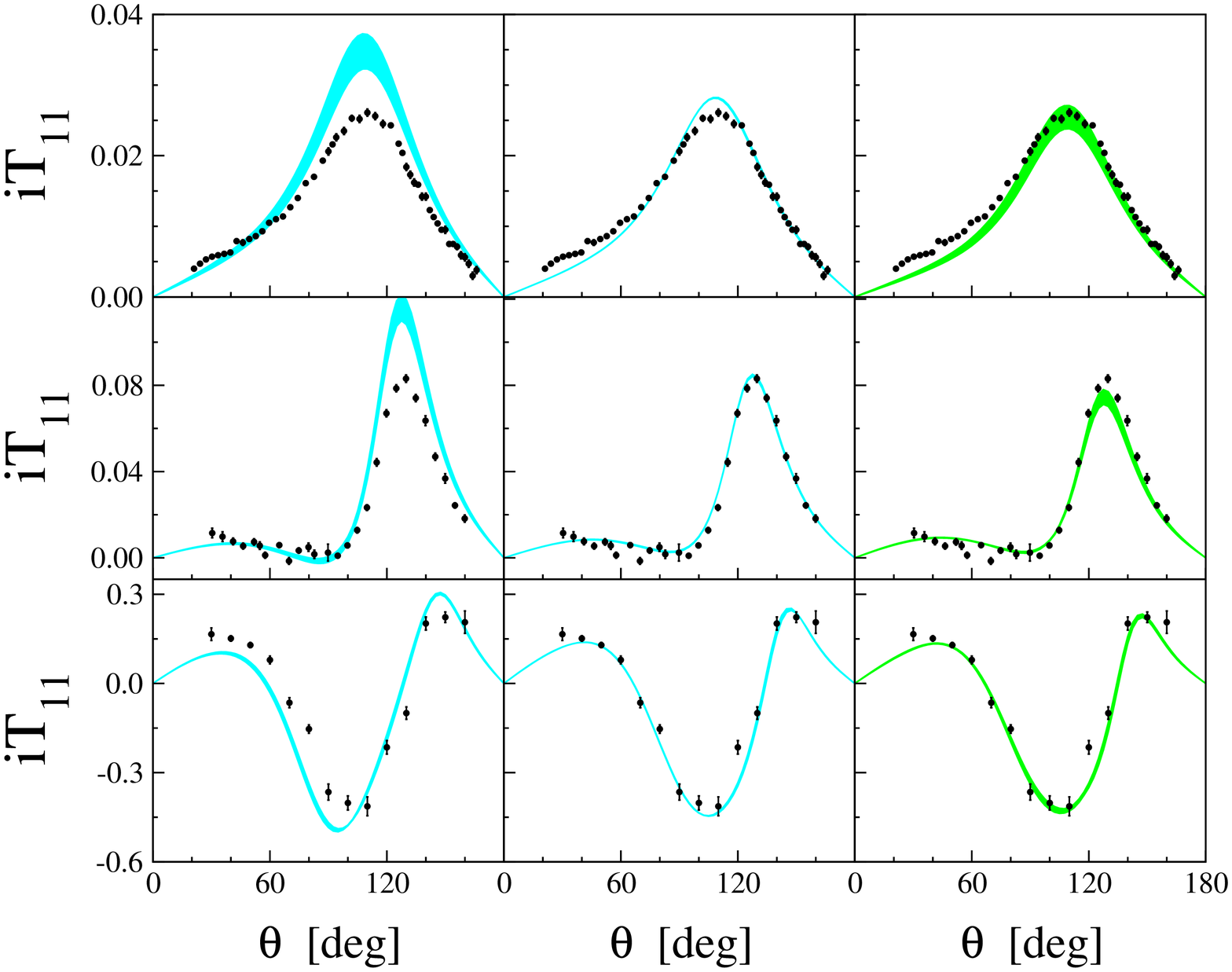}
}
\centerline{\parbox{14cm}{
\caption{\label{fig:t11}Tensor analyzing power $iT_{11}$ for elastic $nd$
scattering, for $E_{\rm lab} = 3,10,65\,$MeV (top to bottom).
The circles are $pd$ data: at 3 MeV from \cite{Shim95}, at 10 MeV from
\cite{Sper84}, and at 65 MeV from \cite{Wit93}.
For further notations, see fig.\ref{fig:sig}.}}}
\end{figure}

\begin{figure}[htb]
\centerline{
\epsfysize=12.5cm
\epsffile{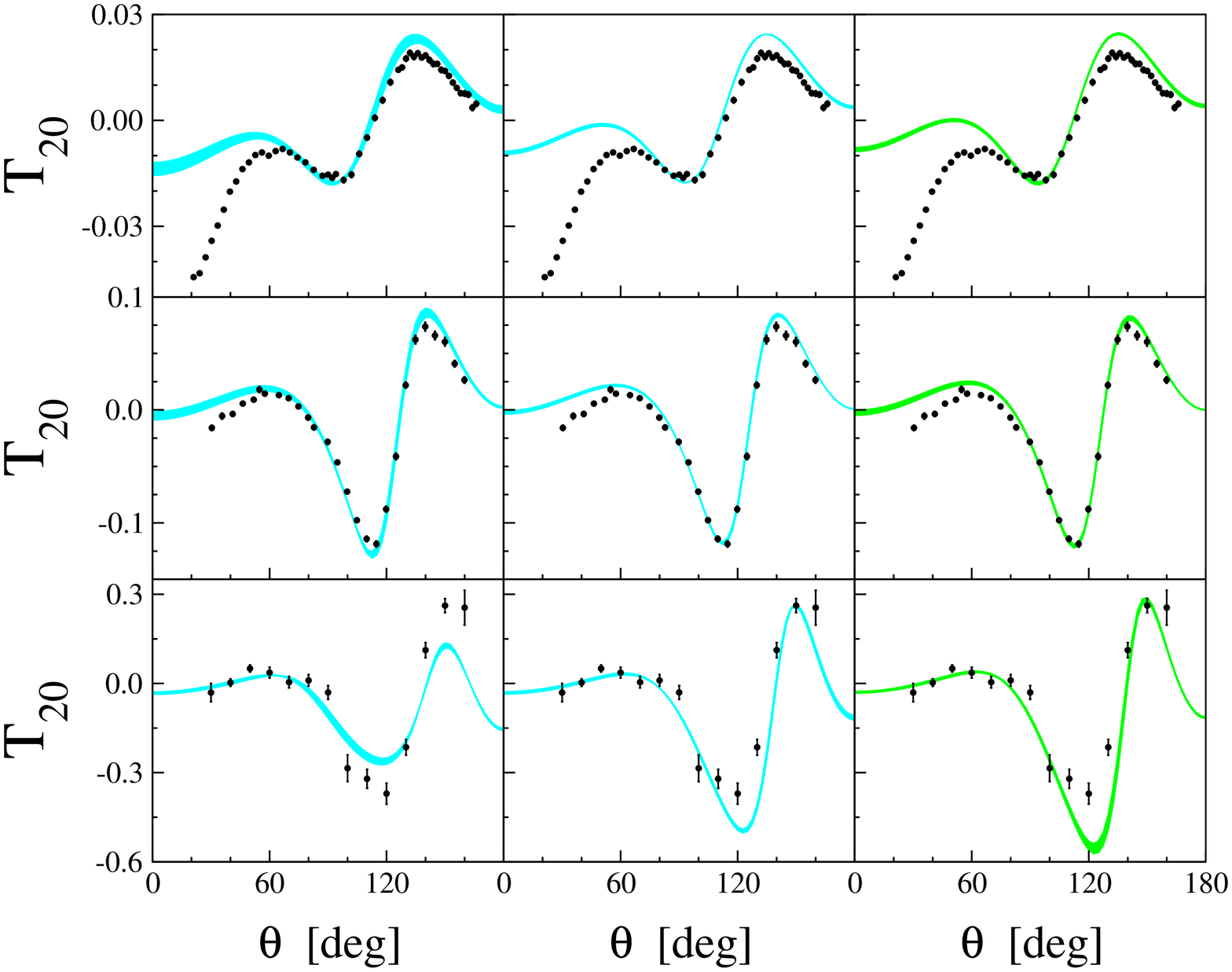}
}
\centerline{\parbox{14cm}{
\caption{\label{fig:t20}Tensor analyzing power $T_{20}$ for elastic $nd$
scattering, for $E_{\rm lab} = 3,10,65\,$MeV (top to bottom).
The circles are $pd$ data: at 3 MeV from \cite{Shim95}, at 10 MeV from \cite{Sper84}, 
and at 65 MeV from \cite{Wit93}.
For further notations, see fig.\ref{fig:sig}.}}}
\end{figure}

\begin{figure}[htb]
\centerline{
\epsfysize=12.5cm
\epsffile{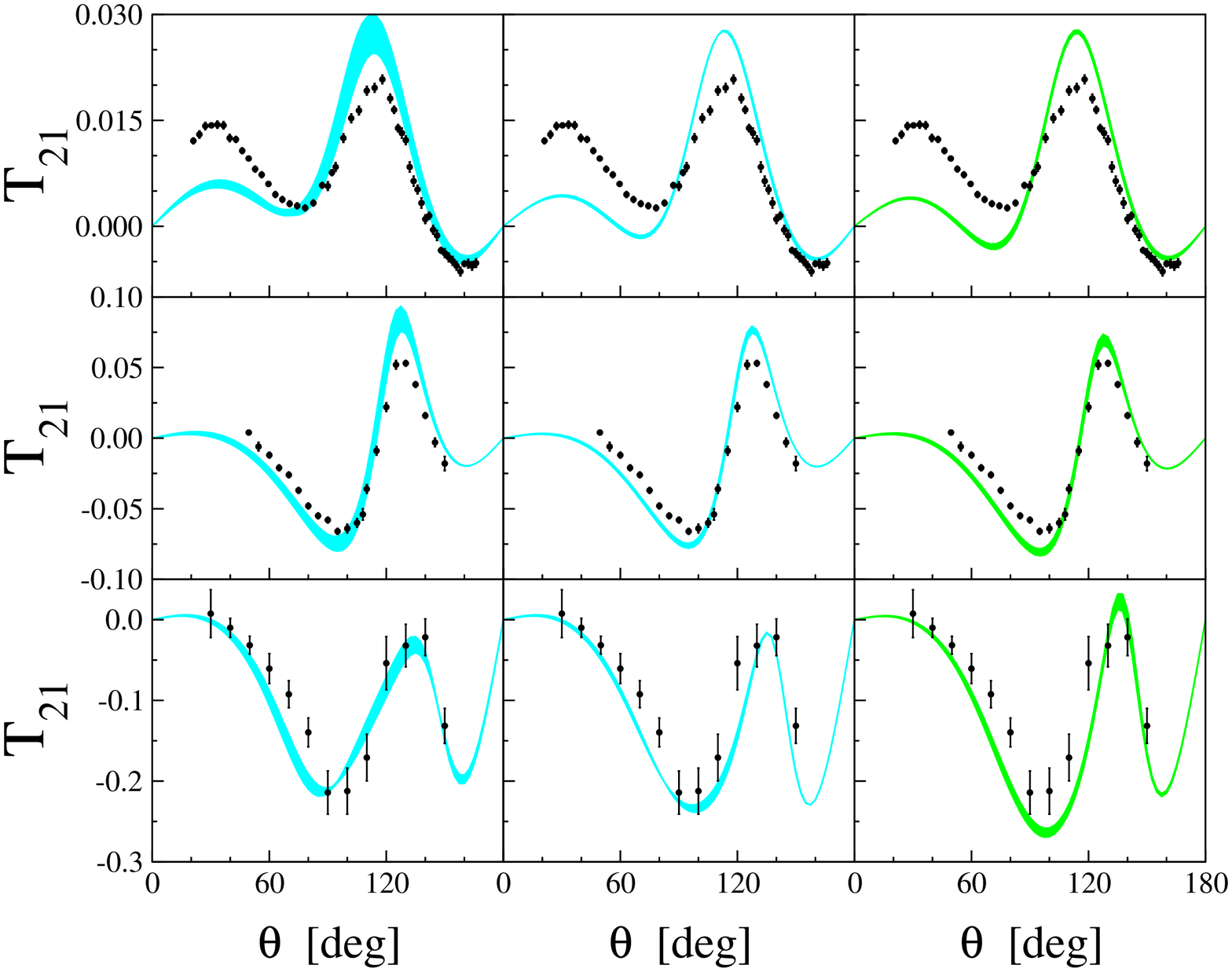}
}
\centerline{\parbox{14cm}{
\caption{\label{fig:t21}Tensor analyzing power $T_{21}$ for elastic $nd$
scattering, for $E_{\rm lab} = 3,10,65\,$MeV (top to bottom).
The circles are $pd$ data: at 3 MeV from \cite{Shim95}, at 10 MeV from \cite{Sper84}, 
and at 65 MeV from \cite{Wit93}.
For further notations, see fig.\ref{fig:sig}.}}}
\end{figure}

\begin{figure}[htb]
\centerline{
\epsfysize=12.5cm
\epsffile{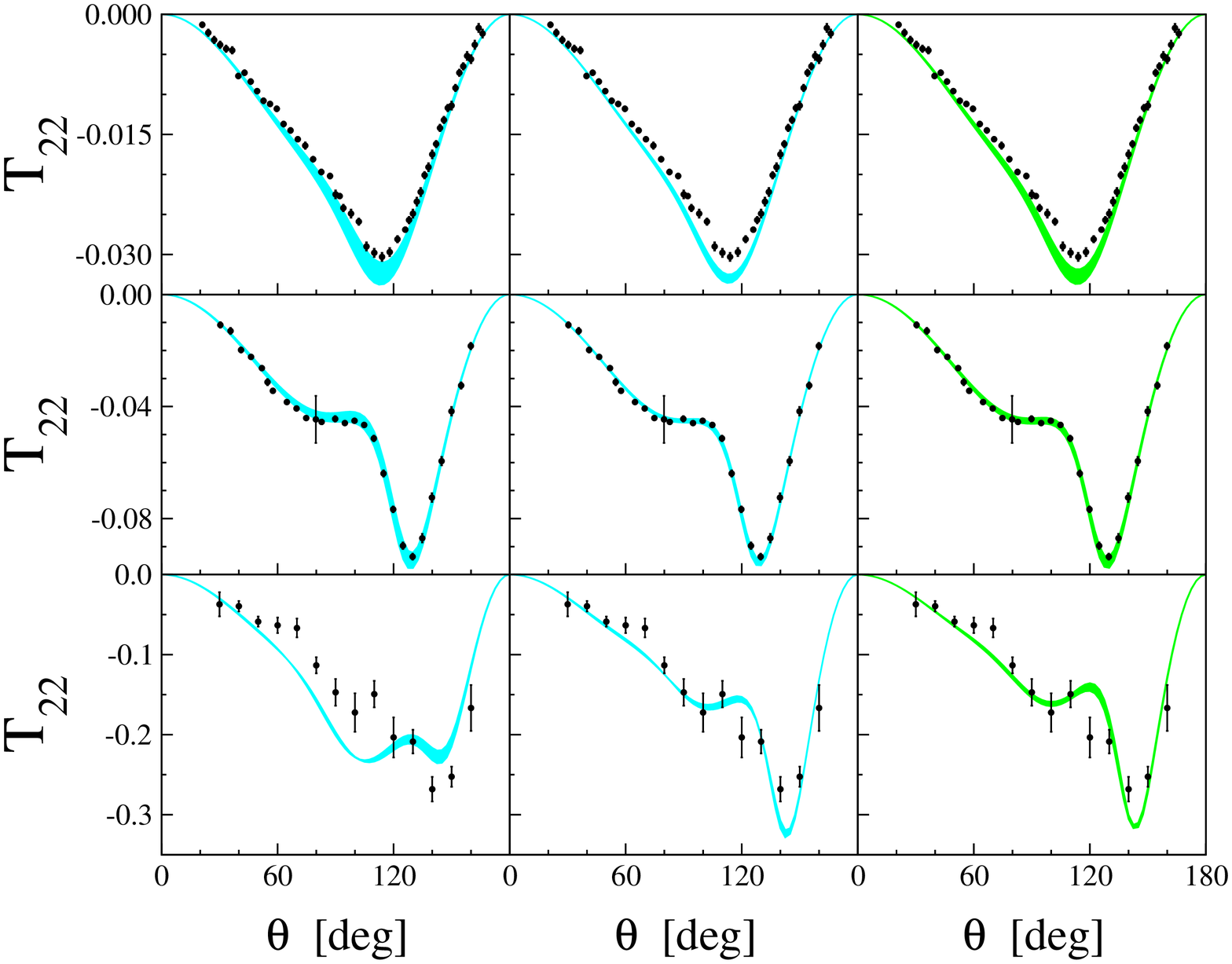}
}
\centerline{\parbox{14cm}{
\caption{\label{fig:t22}Tensor analyzing power $T_{22}$ for elastic $nd$
scattering, for $E_{\rm lab} = 3,10,65\,$MeV (top to bottom).
The circles are $pd$ data: at 3 MeV from \cite{Shim95}, at 10 MeV from \cite{Sper84}, 
and at 65 MeV from \cite{Wit93}.
For further notations, see fig.\ref{fig:sig}.}}}
\end{figure}

\begin{figure}[htb]
\centerline{
\epsfysize=11cm
\epsffile{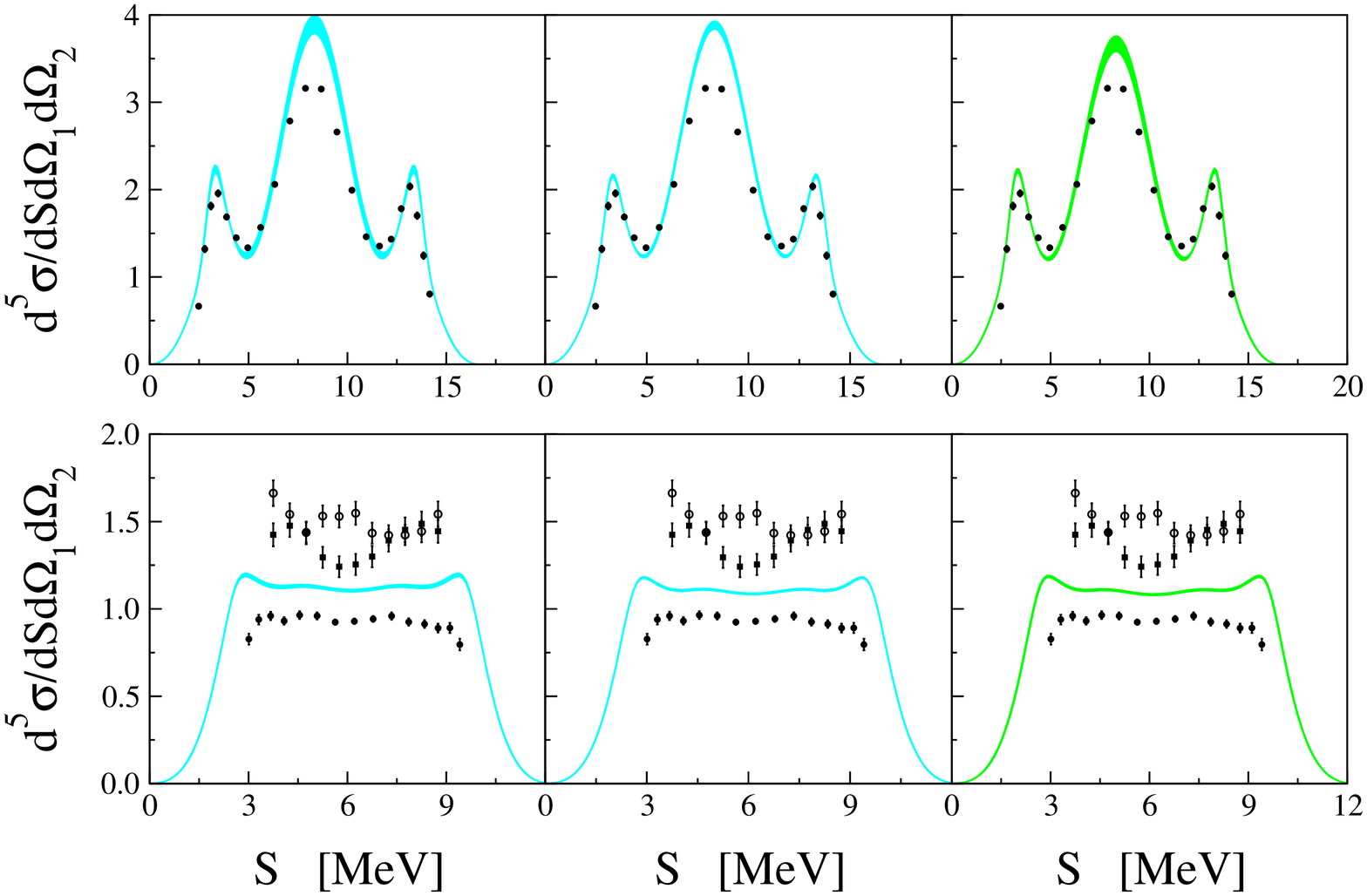}
}
\centerline{\parbox{14cm}{
\caption{\label{fig:breakup}
3N break--up cross sections in [mb/MeV/sr$^2$]  
against the arc length S of the kinematically allowed
locus. The $pd$ data (filled circles) \cite{Raup91} show in the upper row a peak (in the middle)
related to a quasi--free scattering picture. The two sets of $nd$ data from \cite{Set96} 
(open circles) and \cite{Stra88} (filled squares),
upper group, and the $pd$ data, lower ones, are at and in the neighborhood 
of ``space--star'' configuration.}}}
\end{figure}

For 3N scattering we show in  
figs.~\ref{fig:sig}-\ref{fig:t22}
elastic Nd scattering observables for laboratory energies of 3, 10 and 65~MeV, in order,
and in fig.~\ref{fig:breakup} Nd break-up cross sections 
for two arbitrarily selected kinematical configurations
at $E_{\rm lab}=$13 MeV.
In each case
the NLO are compared to the  NNLO* predictions and the ones based on the
modern high--precision potentials. 
Like for the bound state
energies we find in all cases a much reduced cut--off dependence for NNLO* in
comparison to NLO. 
Also, at the highest energy we considered, 65 MeV,  
one observes now a strong improvement compared to the NLO results, which in 
some cases deviate  significantly from the data.  
We also observe that the theoretical uncertainty
due to the cut--off variation is sometimes smaller than the spread using the
various phase equivalent conventional potentials.
Note that most of the deviations of the theoretical predictions 
from the $pd$ data in case of the tensor analyzing
powers and the differential cross section at low energies 
and at forward angles are due to the Coulomb $pp$ force \cite{Kievsky}. 

Let us now take a closer look at the calculated elastic observables.
The differential cross section at NNLO* agrees well with the data  and 
with the predictions based upon various high--precision potentials, cf.  fig.~\ref{fig:sig}, and
is strongly improved at 65 MeV compared to the NLO results.
\renewcommand{\arraystretch}{1.2}    

\begin{table}[htb] 
\begin{center}

\vskip 0.6 true cm 
\begin{tabular}{||l||c|c||c|c||c||c||}
    \hline
     & \multicolumn{2}{|c||}{NLO}  &  \multicolumn{2}{|c||}{NNLO*}  &   &     \\ 
    \cline{2-5}   
 \rb{$T_{\rm lab}$ [MeV]}   & 500 MeV  & 600 MeV & 500 MeV & 600 MeV & \rb{CD-Bonn} & \rb{NPSA} \\
\hline  \hline
$^3P_0$ &  &  &  &  & & \\ 
1    & 0.19 & 0.19  & 0.17 & 0.17 & 0.18  & 0.18 \\
5    & 1.68 & 1.67 & 1.59 & 1.58 & 1.61  & 1.63(1) \\
10   & 3.74 & 3.72 & 3.62 & 3.58 & 3.62  & 3.65(2) \\
25   & 8.28 & 8.22 & 8.26 & 8.16 & 8.10  & 8.13(5) \\
50   & 10.90 & 10.84 & 11.12 & 11.01 & 10.74 & 10.70(9)\\
100  & 8.27 & 8.31 & 8.34 & 8.43 & 8.57  & 8.46(11)\\
150  & 2.52 & 2.52 & 1.88 & 2.15 & 3.72  & 3.69(14)\\
200  & $-$3.70 & $-$4.11 & $-$5.43 & $-$5.23 & $-$1.55 & $-$1.44(17)\\
\hline \hline
$^3P_1$ &  &  &  &  & & \\ 
1    & $-$0.12 & $-$0.12 & $-$0.11 & $-$0.11 & $-$0.11  & $-$0.11 \\
5    & $-$0.99 & $-$0.99 & $-$0.91 & $-$0.92 & $-$0.93  & $-$0.94 \\
10   & $-$2.17 & $-$2.16 & $-$2.02 & $-$2.02 & $-$2.04  & $-$2.06 \\
25   & $-$5.05 & $-$5.03 & $-$4.82 & $-$4.83 & $-$4.81  & $-$4.88(1) \\
50   & $-$8.35 & $-$8.32 & $-$8.22 & $-$8.23 & $-$8.18 & $-$8.25(2)\\
100  & $-$12.61 & $-$12.66 & $-$13.49 & $-$13.47 & $-$13.23  & $-$13.24(3)\\
150  & $-$15.56 & $-$15.94 & $-$18.48 & $-$18.43 & $-$17.51  & $-$17.46(5)\\
200  & $-$17.80 & $-$18.86 & $-$23.64 & $-$23.65 & $-$21.38  & $-$21.30(7)\\
\hline \hline
$^3P_2$ &  &  & &  & & \\ 
1    & 0.02 & 0.02 & 0.02 & 0.02 & 0.02  & 0.02 \\
5    & 0.24 & 0.24 & 0.25 & 0.25 & 0.26  & 0.25 \\
10   & 0.70 & 0.70 & 0.71 & 0.71 & 0.72  & 0.71 \\
25   & 2.87 & 2.89 & 2.64 & 2.65 & 2.60  & 2.56(1) \\
50   & 8.05 & 8.29 & 6.29 & 6.34 & 5.93 & 5.89(2)\\
100  & 20.32 & 22.60 & 11.31 & 11.70 & 11.01  & 10.94(3)\\
150  & 29.73 & 35.97 & 12.11 & 13.04 & 13.98  & 13.84(4)\\
200  & 34.02 & 44.30 & 9.92 & 11.31 & 15.66 & 15.46(5)\\

    \hline
  \end{tabular}
\caption{$^3P_j$ $np$ phase shifts at NLO and NNLO* for the smallest and 
largest values of the cut--off compared to the phases based on the CD-Bonn 
potential \cite{cdb2000} and to the Nijmegen PSA \cite{NPSA}.
\label{tab_pwaves}}
\end{center}
\end{table}
The vector analyzing power of 
elastic $nd$ scattering at low energies is well known to be underpredicted 
by the standard NN potential models, see fig.~\ref{fig:ay}, right panel, and this remains true 
even after inclusion of the existing 3N forces based
on boson exchanges.
As reported in ref.~\cite{ourPRL} and shown in the left panel of fig.~\ref{fig:ay}, 
the NLO predictions at 3 MeV are essentially in agreement with the data, while at 
10 MeV one even observes a slight overestimation in maximum. The NLO results for $A_y$
at 65 MeV show significant deviations from the data. Our predictions at NNLO* are 
much closer to the results based upon the high-precision potentials, i.e. the data are
underpredicted at low energies (3 and 10 MeV) and reproduced accurately at higher ones (65 MeV), 
cf. fig.~\ref{fig:ay}.
Although some improvement with respect to the predictions based upon the high-precision potentials
can be seen at 3 and especially at 10 MeV, the pending puzzle is now back at NNLO*.
As pointed out in ref.~\cite{entem_ay}, one possible reason for the significant change of about 20 \%
in the $A_y$ predictions when going from NLO to NNLO* may be the deviations of the $np$ $^3P_j$
phase shifts from the data at NLO. These channels are well 
known to be very important for the $nd$ $A_y$, see e.g.~\cite{GloeckleREP}.
In table \ref{tab_pwaves} we demonstrate that these partial waves are now much better
described at NNLO*. We also remind the reader that in contrast to high--precision potential
models, which are constructed to perfectly reproduce the NN data below the pion production threshold,
in EFT one does not aim at a perfect description of the data by increasing the phenomenological
content of the NN interaction but rather at performing systematic order--by--order calculations.
At each specific order in the low--energy expansion (in our case chiral expansion) one 
has some theoretical error due to missing higher order terms, which can be estimated. Considering our 
results for $A_y$ at NLO one should therefore keep in mind 
the level of precision of the NLO approximation. Further, since the $nd$ $A_y$ is a 
very sensitive observable and is strongly affected by changing the  $np$ $^3P_j$ phase shifts 
by only few percent, the large uncertainty for this specific observable has to be expected 
in the EFT approach.

The situation with the deuteron vector analyzing power $i T_{11}$ 
is very similar to the one with $A_y$. This is shown 
in fig.~\ref{fig:t11}.\footnote{Notice that only $pd$ data exist for this 
observable. Inclusion of the Coulomb interaction will lead to significant underestimation
of the $i T_{11}$ \cite{Kievsky}.} The NNLO* predictions for the tensor analyzing powers $T_{20}$ and 
$T_{21}$  at 3 and 10 MeV as well as for $T_{22}$ at all three energies follow 
the band made up from the variations among the high--precision potentials. 
Remarkably, our results for $T_{20}$ and $T_{21}$ are even significantly closer to the data at 65 MeV. 

In case of the specific 3N break--up results shown in fig.\ref{fig:breakup} 
the chiral force predictions are equally off the data
as the predictions of the conventional forces. 
In case of the upper row the deviations in the quasi--free peak to the $pd$ data
might be due to Coulomb force effects, whose precise size is still unknown. 
The lower row addresses the space--star anomaly. We underestimate significantly the two sets
of $nd$ data, which are also far off the $pd$ data. 
As in the case of elastic scattering observables the NNLO* predictions follow the 
band made up from the various high--precision potentials.
Again the size of Coulomb force effects is
unknown. For more information on these break--up configurations see refs.~\cite{GloeckleREP},
\cite{Set96}.

It is also interesting to compare our results to the ones shown in ref.~\cite{entem_ay},
in which the same $nd$ scattering observables have been calculated using the phenomenological
high--precision 
extension of the chiral potential by Entem and Machleidt \cite{entem}. In fact, our results
for these observables 
show a remarkable similarity to the ones presented in this reference, i.e. both predictions
agree with the calculations based upon the conventional high--precision 
potential models and with the data in most cases and are slightly closer to the data for  
$T_{20}$ and $T_{21}$ at 65 MeV. The only significant differences between our results and
the ones of ref.~\cite{entem_ay} are observed for $A_y$ (and $i T_{11}$) at low
energies (3 and 10 MeV), which are slightly improved in case of the NNLO* version. 
It is very gratifying  to see that at least up to $E_{\rm lab} = 65$ MeV our NNLO*
potential with 11 adjustable parameters works for $nd$ scattering equally 
well as the one of ref.\cite{entem}
with 46 adjustable parameters. 
This remarkable agreement may serve as a nice demonstration of the power and the advantage 
of an EFT with consistent power counting
compared to more phenomenological approaches: performing chiral expansion 
of the nuclear force up to some definite order by inclusion of {\bf all} relevant 
diagrams and counter terms allows to describe low--energy observables with the same 
precision regardless of the kind of system the theory is applied to (2N, 3N, ...). 
From the point of view of EFT, it makes not much sense to improve the description of the
2N observables alone by a phenomenological extension of the short--range part of the NN force. 
As one can see comparing figs.\ref{fig:sig}-\ref{fig:t22} with the corresponding ones
of ref.~\cite{entem_ay}, this does not lead to an improvement in describing other systems 
at low energy (i.e. the 3N system). In order to reduce the theoretical uncertainty, one 
should instead go to higher orders, which requires the inclusion of 3N, 4N, ..., interactions
as well as more pion exchanges in the 2N force. 
Furthermore, the whole concept of developing phenomenological NN potentials, which 
reproduce the NN data perfectly with $\chi^2/$datum$=1$ is in conflict with the 
general EFT philosophy:
at each fixed finite order of the low--energy expansion one necessarily has some 
definite uncertainty in description of observables. Adjusting the cut--off parameters 
in various partial waves to improve the fit to data, as it has been done in ref.\cite{entem}, 
is not acceptable from the point of view of pure EFT, 
where the cut--off dependence of observables may serve as 
an estimation of the theoretical error.  


It is now an urgent task to encode the three topologically different 3N
forces, which have to be taken into account  at  NNLO (NNLO*)
and to determine the corresponding 
parameters in the 3N system. Pioneering studies in \cite{Hueber} indicate that
specifically the diagram in fig. [5] of this reference 
might have a chance to solve the $A_y$ puzzle. 
This extensive work will be dealt with in  a forthcoming paper.

\section{Summary}
\setcounter{equation}{0}

The concept and the resulting NN forces at LO, NLO and NNLO of $\chi$PT
have been reviewed. Our approach is
based on the method of unitary transformation applied to the most
general chirally invariant
Hamiltonian expressed in terms of pion and nucleon fields. This method
leads to energy independent
nuclear forces, a property which is important for the application to 
more than two nucleon systems.
The NNLO NN forces driven by the low energy constants $c_{1,3,4}$ 
lead to
deeply bound unphysical NN
states in low partial waves if the values $c_{1,3,4}$ are taken from typical
$\pi$N data analysis.\footnote{Note that this statement might not hold true for
different regularization schemes.}
While
this has no negative observable consequences in the NN system, since the spurious NN
bound state energies are
outside the realm of validity of $\chi$PT, they lead to a  scenario  for nuclear
physics which is quite different from the one driven by conventional 
nuclear forces. First, the central part of the NN potential turns out
to be much more attractive as it is expected from conventional approaches.
Further, the predictions for 3N, 4N, ..., binding energies 
based upon the purely NN forces 
are much lower, far below the experimental
values, and 3N scattering observables deviate dramatically from the data.
Therefore, unlike for
conventional NN forces, which to a very large extent describe the data,
and 3N forces are only
needed as a relatively small additional contribution, the 3N force contributions
here  will be very essential. We provided arguments based upon experiences with 
meson theoretical potentials supporting the
choice of $c_{3,4}$ constants, which are numerically smaller and where
intermediate 
$\Delta$--contributions are subtracted out. Based on those values we introduced
a novel NNLO* NN force which
describes NN phase shifts with comparable quality as the NNLO one up
to about $E_{\rm lab} = 200$ MeV.
These NNLO* potential is free of spurious bound states and leads to
predictions in the 3N and 4N
systems which are rather close to the ones familiar from conventional
high precision NN forces. It is
now of highest interest to include the 3N forces which should be taken
into account at that order in
$\chi$PT. This work is in preparation.

In contrast to conventional nuclear forces this chiral approach is
systematic in the sense of power counting and nuclear forces are expected
to be constructed in a
convergent scheme. Therefore the step to NNNLO should be performed in 
order to see whether convergence
can be reached and long pending problems with conventional forces
like the low energy analyzing
power $A_y$ can be solved without ad hoc assumptions.

\section*{Acknowledgments}
\setcounter{equation}{0}
We would like to thank Joanna 
Kur\'{o}s-{\.Z}o{\l}nierczuk 
for helping us in creating various figures.
This work was supported by the Deutsche Forschungsgemeinschaft (E.E.), the U.S. National Science Foundation 
under Grant No.~PHY-0070858
(A.N.) and the Polish Committee for Scientific Research under Grant No.~2P03B02818 (H.W.). 
The calculations have been performed on the T90 and the T3E of the NIC J\"ulich,
Germany.

\end{document}